\documentclass[a4paper]{JHEP3}%
\usepackage{amsmath}%
\usepackage{epsfig}%
\usepackage{cite}%
\newcommand{\hh}{\tilde{h}}%
\newcommand{\ff}{\tilde{f}}%
\def\hhh{\Tilde{\Tilde{h}}}%
\def\fff{\Tilde{\Tilde{f}}}%

\title{The fate of the zero mode of the five-dimensional kink in the presence of gravity}

\author{M. Shaposhnikov\\ 
Institut de Th\'eorie des Ph\'enom\`enes Physiques, 
Ecole Polytechnique F\'ed\'erale de Lausanne, CH-1015 Lausanne, Switzerland\\ 
E-mail: \email{Mikhail.Shaposhnikov@epfl.ch}}

\author{P. Tinyakov\\
Service de Physique Th\'eorique, Universit\'e Libre de Bruxelles, 
CP225, bld.~du Triomphe, B-1050 Bruxelles, Belgium\\
E-mail: \email{Petr.Tiniakov@ulb.ac.be}}

\author{K. Zuleta\\
Institut de Th\'eorie des Ph\'enom\`enes Physiques, 
Ecole Polytechnique F\'ed\'erale de Lausanne, CH-1015 Lausanne, Switzerland\\ 
E-mail: \email{Katarzyna.ZuletaEstrugo@epfl.ch}}

\abstract{We investigate what becomes of the translational zero-mode of a five-dimensional domain wall in the presence of gravity, studying the scalar perturbations of a thick gravitating domain wall with~$AdS$ asymptotics and a well-defined zero-gravity limit. Our analysis reveals the presence of a wide resonance which can be seen as a remnant of the translational zero-mode present in the domain wall in the absence of gravity and which ensures a continuous change of the physical quantities (such as e.g. static potential between sources) when the Planck mass is sent to infinity. Provided that the thickness of the wall is much smaller than the $AdS$ radius of the space-time, the parameters of this resonance do not depend on details of the domain wall's structure, but solely on the geometry of the space-time.} 


\preprint{}

\begin{document}

\section{Introduction} 
 
The proposal that our universe could be a topological defect in a higher-dimensional space-time, which is at the basis of the brane-world scenarios, was put forward already in the~80's~\cite{shap,akama} as an alternative to Kaluza-Klein type theories~\cite{Kaluza,Klein}. Contrarily to the Kaluza-Klein type setups, in the brane-world models the extra dimensions can be large and even infinite. The key idea is that by trapping the matter fields on the defect (a domain wall or a string) one can obtain a low-energy effective theory which looks four-dimensional, the extra dimensions becoming visible only in the very high energy experiments. The zero modes of the higher-dimensional fields, confined on the defect, can be interpreted as the four-dimensional particles. In the setup of Ref.~\cite{shap} the localization was achieved for two types of fields. Firstly, among the excitations of the scalar field forming the wall there was a localized zero-mode, which could be seen as a massless~4D scalar particle. Secondly, the model naturally incorporated fermions, allowing trapping of the fermion zero mode via Yukawa interaction between the fermion field and the domain wall.

The interest in higher-dimensional theories of this type increased tremendously few years ago, when it was discovered
that this kind of picture could in fact assimilate gravity. 

Firstly, in~\cite{add,add_2} a model was proposed where our universe is modeled by a thin~\mbox{3-brane} embedded in a higher-dimensional space-time with compact extra dimensions; the influence of the brane's tension on the geometry of the bulk is assumed negligible. All the Standard Model fields are supposed to be trapped on the brane, only the gravity being able to propagate in the bulk. As a consequence, the extra dimensions could be fairly large without contradiction with the available small-distance Newton's law measurements~(at the time when this setup was proposed, even the dimensions of order of $1\,\mathrm{mm}$ were not excluded; the recent experimental bounds~\cite{Hoyle:2004cw} require the extra dimensions to be not larger than~$160\,\mu\mathrm{m}$). The spectrum of perturbations in the models of this type has been studied extensively~\cite{add_3,kugo_1,kugo_2,Giudice:2000av}, and it was found to contain Kaluza-Klein modes of gravitons, as well as exotic scalar particles, called branons~\cite{branon_dyn}. Branons can be identified as the Goldstone bosons associated with the spontaneous breaking of the isometries of the bulk, caused by the presence of the brane. Massive branons were advocated as~dark matter candidates~\cite{dark_branons}.
 
Soon after the proposal of~\cite{add,add_2}, the authors of~\cite{rs} have built two brane world models: firstly a compact one, involving two delta-like 3-branes located at the fixed points of an orbifolded $AdS_5$ space-time~(RS1), and subsequently a non-compact single brane model~(RS2), obtained by letting the compactification radius in~RS1 to be infinite. Considering the graviton spectrum in these scenarios, they have shown that it was possible to reconcile large (or even infinite) extra dimensions with Newtonian gravity. The effective four-dimensional spectrum of both the above models was studied in detail by several authors (see e.g.~\cite{Boos:2002,Rattazzi:loops}). It was found that in the~RS1 model, the effective theory contains a massless spin-two tensor field (four-dimensional graviton), an infinite tower of the massive tensor fields (Kaluza-Klein gravitons), and a massless scalar field called radion. When sending the size of the extra dimension to infinity, thus passing to the non-compact~RS2 model, the radion can be gauged away and disappears from the effective theory~\cite{Boos:2002,Rattazzi:loops,Rattazzi:radion,Giudice:2000av}. The spectrum of the~RS2 model consists of a localized zero mode and a continuum of massive modes. Even if there is no gap separating the zero mode from the continuum, the law of gravity on the wall is essentially Newtonian, the corrections yielded by the light continuum modes modifying this behavior only at short distances. 
 
The infinitely thin branes used in~\cite{rs} may be thought as mere approximations of smooth structures, thick branes, obtainable as solutions to coupled gravitational and matter equations. Regularized versions of the~RS2 model were constructed~\cite{kehagias, kehagias_2, gremm, massimo} along the same lines as in~Ref.~\cite{shap}: smooth gravitating domain walls, generating warped geometry, can be produced by a scalar field~$\Phi$ with a potential spontaneously breaking the~$\Phi\to-\Phi$ symmetry. The spectrum of fluctuations in the regular warped backgrounds has also been studied, e.g. in~\cite{massimo} (generalizing Bardeen's formalism of gauge invariant variables) and in~\cite{daemi_shap} using the light cone gauge formalism. In this case as well, the only localized particle was the massless spin-two tensor field. 

What may seem surprising in the results concerning the RS2 model and its smooth generalizations, is that they do not show a slightest trace of the massless scalar present in the setup of Ref.~\cite{shap}. It seems that in the process of including gravity it has somehow ``evaporated''. However, it is rather strange that a particle would disappear altogether as soon as the gravity is switched on, no matter how weakly. Locally, a gravitating domain wall should resemble a non-gravitating one, that is the model of Ref.~\cite{shap}. Presence of a massless particle is bound to have physical consequences -- for example we expect it to induce a long-range potential between sources located on the brane. The physical quantities (such as, for example, static potential between sources) must change in a continuous manner when the Planck mass is sent to infinity. There should therefore be some imprint of the translational zero mode of~\cite{shap} on the spectrum of scalar perturbations in the presence of gravity, the simplest possibility being that it acquires a finite width or, in other words, becomes a resonance -- either a standard, massive one~\cite{drt}, or maybe a massless quasilocalized mode, of a kind encountered e.g. in the GRS model~\cite{grs}.

In this paper we investigate the fate of the scalar zero mode in the presence of gravity, considering a thick scalar brane model introduced in~\cite{kehagias} which possesses both an~$AdS_5$ asymptotics and a well defined zero-gravity limit. We use the quadratic action for the scalar perturbations derived in~\cite{daemi_shap} to study the interactions of these perturbations with sources localized on the brane. Our analysis shows that although there aren't any localized modes or even narrow resonances, the light modes of the continuous spectrum ensure the continuity of the physical quantities. From the point of view of the four-dimensional physics, the behavior of these modes can be interpreted as a wide resonance, whose mass and width are entirely determined by the geometry of the space-time and do not depend on details of the wall's structure, provided that the wall thickness is much smaller than the $AdS$ radius~$R_{\mathrm{AdS}}$ of the space-time. The influence of these modes is significant at distances much smaller than~$R_{\mathrm{AdS}}$; when the distances~$R$ are much larger than~$R_{\mathrm{AdS}}$, the potential acquires a~$1/R^3$ form and its amplitude is strongly suppressed. We find similar results studying propagation of waves emitted by an oscillating source: for the waves with frequencies larger than the inverse~$AdS$ radius the standard~$1/R$ dependence of the amplitude on the distance to the source is recovered in a large interval of distances.

The paper is organized as follows: In section~\ref{sect-domain}, we remind the construction of the thick domain walls -- first in the absence and next in the presence of gravity. Section~\ref{sect-scalar} contains the analysis of the spectrum of the system of scalar perturbations of a gravitating domain wall. Sections~\ref{sect-static} and~\ref{sect-periodic} are dedicated to the physical applications: in section~\ref{sect-static}, we determine the static potential between two sources mediated by the scalar perturbations and in the section~\ref{sect-periodic} we generalize the study to the case of the moving sources. 

\section{\label{sect-domain}Domain walls in 5D space-times}

\subsection{Non-gravitating case -- domain wall in the flat space-time}

To begin, let us remind in more detail the results of~\cite{shap}, where an explicit field theoretical realization of the mechanism of localization of matter on a thick domain wall (without gravity) was devised. The main idea of~\cite{shap} is to break the translation invariance along the extra dimension by forming a domain wall on which the particles will be trapped. This can be achieved by a quantum field model containing a real scalar field with the double-well potential. The corresponding action is:
\begin{equation}
\label{action_phi}
S=\int d^4x\,dr \left[ \frac{1}{2}\eta^{MN}\partial_M\Phi\partial_N\Phi -\frac{\lambda}{4}(\Phi^2-v^2)^2\right]\ . 
\end{equation}
The classical equations of motion admit a domain wall solution $\Phi_c(r)$ interpolating between the two classical vacua~$\Phi_{\mathrm{vac}}=\pm v$, whose form coincides with the one-dimensional kink:
\begin{equation}
\label{kink}
\Phi_c(r)=v\tanh(ar) \ ,
\end{equation}
where $a^2=\lambda v^2 /2$. The perturbations around this background~$\phi=\Phi-\Phi_c$ satisfy the linearized equation of motion:
\begin{equation}
\label{phi_perts_eqn}
\partial^M\partial_M\phi(x,r)+\lambda(3\Phi^2_c(r)-v^2)\phi(x,r)=0 
\end{equation}
and therefore can be written as 
$$
\phi(x,r)= \sum_k \hspace{-0.5cm} \int u_k(x)\eta_k(r)\ ,
$$ 
where $u_k(x)$ satisfy the usual four-dimensional Klein-Gordon equation
$$
\left[\partial^\mu\partial_\mu +\mu^2\right]u_k(x)=0\ ,
$$
with mass $\mu^2=k^\nu k_\nu$ and where $\eta_k(r)$ are the normal modes of the one-dimensional kink:
\begin{equation}
\label{kink_wave_f}
-\partial^2_r\eta_k(r)+\lambda(3\Phi^2_c(r)-v^2)\eta_k(r)= \mu^2 \eta_k(r)\ .
\end{equation}
The spectrum of~$\phi$ comprises a zero-mode~($\mu^2=0$):
\begin{equation}
\label{phi_zero_mode}
\phi_0(x,r)=\frac{\sqrt{3a}}{2}\frac{1}{\cosh^2(ar)}u_0(x) \ ,
\end{equation}
which can be interpreted as a massless scalar particle localized on the wall. There is also a localized heavy mode with mass $\mu^2=3a^2$ and the continuous spectrum starting at~$\mu^2=4 a^2$, corresponding to the perturbations which are not confined inside the wall.

For the sake of completeness, let us add that the model can naturally account for the massless four-dimensional fermions, localizing the zero mode of the bulk fermion through the Yukawa interaction with the domain wall.

In this work, it is the massless scalar mode~$\phi_0(x,r)$ and the low energy physics related with it which will be the focus of our attention. We expect the particles of the type~\eqref{phi_zero_mode} to behave as fairly ordinary four dimensional particles: collide, form bound states, etc. Given that the particles~$\phi_0(x,r)$ are massless we expect them to mediate long-range potential between sources on the domain wall.

\subsection{Background solution for a self-gravitating domain wall}

Let us now consider again a theory of a real scalar field~$\Phi$ in a five-dimensional space-time, but this time adding an ingredient which was absent in the considerations of Ref.~\cite{shap}: the gravity. Supposing minimal coupling of the scalar to gravity we obtain the action:
\begin{equation}
\label{action_grav_kink}
S=\int d^4x\,dr \sqrt{g}\left[-\frac{R}{2\kappa}+\frac{1}{2}g^{MN}\partial_M\Phi\partial_N\Phi -U(\Phi)\right]\ , 
\end{equation}
where~$\kappa$ denotes the five-dimensional gravitational constant.
In analogy to the zero gravity case described in the previous section, we suppose that the potential satisfies~$U(\Phi)=U(-\Phi)$ and that the field~$\Phi$ is in a domain wall configuration. The presence of the scalar field determines the geometry of the space-time via the Einstein equations
\begin{equation}
\label{Einstein}
R_{MN}-\frac{1}{2}g_{MN}R=\kappa\left[\partial_M\Phi\partial_N\Phi-g_{MN}\left(\frac{1}{2}g^{AB}\partial_A\Phi\partial_B\Phi -U(\Phi)\right)\right]
\end{equation}
and the curvature of the space-time modifies the field equation of the scalar:
\begin{equation}
\label{phi_motion}
\frac{1}{\sqrt{g}}\,\partial_M\left[\sqrt{g}\,g^{MN}\partial_N\Phi\right]+\frac{\partial U}{\partial\Phi} =0\ . 
\end{equation}
In order to find the background configuration of the model, one therefore has to solve the coupled system of equations~\eqref{Einstein} and~\eqref{phi_motion}. We suppose that the space-time possesses a warped geometry, with the metric of the form:
\begin{equation}
\label{metric}
ds^2=g_{MN}dx^Mdx^N=e^{A(r)}\eta_{\mu\nu}dx^\mu dx^\nu -dr^2 \ ,
\end{equation}
the warp factor $e^{A(r)}$ being a smooth function of the extra coordinate~$r$, with $A(r)$ symmetric and decreasing. The four-dimensional Planck mass is now finite and is defined by the relation:
$$
M_4^2\equiv M_5^3\int_{-\infty}^\infty\ dr \, e^{A(r)}\ ,
$$
where $M_5=(4\kappa)^{-1/3}$ is the five-dimensional Planck scale. For the metric~\eqref{metric} the background equations~\eqref{Einstein} and~\eqref{phi_motion} read:
\begin{align}\label{bg-eqs}
& \Phi'' + 2A' \Phi' -\frac{\partial U}{\partial\Phi} =0,\nonumber \\ 
& A'^2 = \frac{2\kappa}{3}\left[ \frac12\Phi'^2 - U\right], \\
& A'^2 +A'' = - \frac{2\kappa}{3} \left[\frac{1}{2}\Phi'^2 + U\right].\nonumber
\end{align}
An interesting exact solution for a thick brane background has been found~in Ref.~\cite{kehagias}. In this solution, the scalar field is supposed to be in the familiar kink configuration:
\begin{equation}
\label{Phi}
\Phi(r)=v\tanh(ar)\ ,
\end{equation}
which allows to integrate the background equations~\eqref{bg-eqs}, thus obtaining the exponent of the warp factor of the form:
\begin{equation}
\label{A(r)}
A(r)=-2\beta\ln\cosh^2(ar)-\beta\tanh^2(ar) 
\end{equation}
and the potential:
\begin{equation}
\label{U}
U(\Phi)=\frac{\lambda}{4}(\Phi^2-v^2)^2-\frac{\beta\lambda}{3v^2}\Phi^2(\Phi^2-3v^2)^2 \ ,
\end{equation}
where again $a^2=\lambda v^2 /2$ and where a dimensionless constant $\beta\equiv{\kappa v^2}/{9}$ has been introduced.\footnote{Strictly speaking, the potential~\eqref{U} is not bounded from below. However, we are interested in small perturbations around the kink configuration and therefore will not be concerned by this instability.}
 
Besides being an exact solution of~\eqref{bg-eqs}, the background~\eqref{Phi}--\eqref{U} has two attractive features:

Firstly, it has a well-defined flat limit, tending smoothly to domain wall solution of~\cite{shap} as $\beta$ goes to zero:
\begin{eqnarray*}
A(r)&\longrightarrow & 0 \ ,\\
U(\Phi)&\longrightarrow & \frac{\lambda}{4}(\Phi^2-v^2)^2 \ .
\end{eqnarray*}
It can therefore be considered a self-gravitating extension of~\cite{shap}, which makes it a perfect background to investigate the fate of the zero mode of the five-dimensional kink in the presence of gravity. 

Secondly, the solution~\eqref{Phi}--\eqref{U} can be seen as a regularized version of the~RS2 model~\cite{rs} and there exists a well-defined ``brane'' limit (or, in other words, thin wall limit), in which it becomes~RS2. Indeed, the action~\eqref{action_grav_kink} can be split into two parts: a localized brane part and a bulk part, 
\begin{align*}
S_\mathrm{brane}&=\int d^4x\,dr \sqrt{g}\left[\frac{1}{2}g^{MN}\partial_M\Phi\partial_N\Phi -U(\Phi)+U_\infty\right]=-\int d^4x \,\sigma\ , \\ 
S_\mathrm{bulk}&=\int d^4x\,dr \sqrt{g}\left[-\frac{R}{2\kappa}-\frac{\Lambda_5}{\kappa}\right]\ ,
\end{align*}
with the effective bulk cosmological constant~$\Lambda_5$ determined by the asymptotics of the potential $U_\infty\equiv\lim_{|r|\to\infty} U(\Phi(r))$: 
$$
\Lambda_5=\kappa U_\infty =-24\beta^2 a^2\ 
$$
and the tension of the brane given by
$$
\sigma=\int dr\sqrt{g}\left[\frac{1}{2}(\partial_r\Phi)^2 +U(\Phi)-U_\infty\right] \ .
$$
In the thin wall limit, which is defined as~\cite{kehagias}:
\begin{equation}
\label{tw}
a\to\infty\ , \qquad \xi^2\equiv a\beta<\infty\ ,
\end{equation}
the warp factor acquires the characteristic~$AdS$ form~$e^{A(r)} \to e^{-4a\beta|r|}$ and the domain wall becomes a delta-like brane, whose tension is tuned to the bulk cosmological constant via the relation~$\sigma^2=-6\Lambda_5/\kappa^2$ familiar from the~RS models.

\section{\label{sect-scalar}Scalar perturbations}

Let us now consider perturbations around the background solution~\eqref{bg-eqs}. When the gravity is switched on, we have perturbations coming both from the metric and from the scalar field. Their equations of motion can be obtained by setting:
\begin{eqnarray*}
g_{MN} & \longrightarrow &g_{MN}+h_{MN}\ , \\
\Phi & \longrightarrow & \Phi +\phi\ , 
\end{eqnarray*}
in the field equations~\eqref{Einstein} and~\eqref{phi_motion} and keeping only terms linear in perturbations. The study of the perturbations becomes now fairly involved, as the perturbations of the scalar field couple to the metric fluctuations. Yet another difficulty comes from the gauge invariance -- not all the degrees of freedom are actually physical. This inconvenience can be dealt with either by constructing gauge invariant variables~\cite{massimo} or by fixing the gauge~\cite{daemi_shap,massimo}. In this work we choose the latter option and use the results of~Ref.~\cite{daemi_shap} where the perturbations were studied in the light cone gauge.

\subsection{General setup} 

In Ref.~\cite{daemi_shap}, a formalism was developed to analyze the spectrum of small perturbations of a general system of gravitational, gauge and scalar fields in~$D$-dimensional space-times (and, in particular, in warped geometries). The bilinear action of the system of perturbations was decoupled there into spin-two, -one and -zero sectors using the light cone gauge. In the light cone coordinates, defined by~$x^M=(x^+,\vec{x}_\perp,x^-,r)=(x^+,x^1,x^2,x^-,r)$, with~$x^\pm=\left(x^0\pm x^3\right)/\sqrt{2}$, the background metric~\eqref{metric} becomes:
$$
ds^2=e^{A(r)}\left[2dx^+dx^- - (dx^1)^2-(dx^2)^2\right]-dr^2 \ .
$$
The light cone gauge corresponds to setting:
$$
h_{-M}=0\qquad \mbox{for all~$M$} \ .
$$
In this gauge the fields of different spins are readily separated from each other. The physical fields are~$h_{12}$ and~$h_{11}-h_{22}$ (gravity multiplet, two degrees of freedom), $h_{ri}$ (vector multiplet or graviphoton, two degrees of freedom) and two scalar fields~$\phi$ and $h_{rr}=-h^i_i$.

We will be interested exclusively in the scalar sector of perturbations. It was shown in~\cite{daemi_shap} that this sector can be reduced to two coupled scalar fields $\chi(x,r)=\sqrt{3/8\kappa}\, h_{rr}$ and~$\phi(x,r)$, where~$\phi$ is the fluctuation of the scalar and~$\chi$ a metric perturbation. For~$D=5$ their action reads:
\begin{align}
\delta S^{(2)}_{\mathrm{scal}}=
\int d^4x\,dr \, e^{2A(r)}\Biggl\{&\frac{1}{2}\left[e^{-A}\eta^{\mu\nu}\partial_\mu \phi\partial_\nu \phi -\phi'^2-\left(2\kappa\Phi'^2+\frac{\partial^2U}{\partial\Phi^2}\right)\phi^2\right]\nonumber \\[.5ex]
+{}&\frac{1}{2} \left[e^{-A}\eta^{\mu\nu}\partial_\mu \chi\partial_\nu \chi -\chi'^2+\left(2A''+A'^2\right)\chi^2\right]\nonumber \\[.5ex]
-{}&\sqrt{\frac{2\kappa}{3}}\left(2\Phi''+A'\Phi'\right)\phi \chi \vphantom{\frac{\partial^2U}{\partial\Phi^2}}\Biggr\} \ .
\label{action}
\end{align}
The fields~$\chi$ and~$\phi$ can be decomposed into four-dimensional plane waves:
$$
\chi\propto \exp(ip_\mu x^\mu) h_p(r)\ , \qquad 
\phi\propto \exp(ip_\mu x^\mu) f_p(r) \ .
$$
The four-dimensional scalar spectrum is therefore determined by the following system of equations:
\begin{equation}
\label{system}
\begin{cases}
\displaystyle \hfill{}-h'' - 2A'h'-\left(2A''+A'^2\right) h+\sqrt{\frac{2\kappa}{3}}\left(2\Phi''+\Phi'A'\right)f=\mu^2 e^{-A} h \ ,\\[1.5ex]
\displaystyle -f '' - 2A'f'+\left(\frac{\Phi'''}{\Phi'}+2A'\frac{\Phi''}{\Phi'}-A''\right)f+\sqrt{\frac{2\kappa}{3}}\left(2\Phi''+\Phi'A'\right)h =\mu^2 e^{-A}f \ ,
\end{cases}
\end{equation}
where $\mu^2=p^\mu p_\mu$ and where we have used the background equation~\eqref{bg-eqs} to express $\frac{\partial^2U}{\partial\Phi^2}$ in terms of $\Phi$ and $A$. For the sake of clarity, we assume that~$A(r)$ and~$\Phi$ are given by the background solution~\eqref{Phi}--\eqref{A(r)}.
 
One should bear in mind that in the light cone formalism the five components of the four-dimensional massive graviton are distributed among all three spin sectors: two of them are in the tensor sector, another two in the vector sector and the remaining component is hidden in the scalar sector. As a consequence, in the scalar sector of perturbations a variable can be found which belongs to the graviton multiplet and which therefore has the same spectrum as the tensor fluctuations. This variable, which can be written as a particular combination of the fields~$\chi$,~$\phi$ and their derivatives, was identified and described in~\cite{daemi_shap}. The main concern of the present work is the physics of the genuine four-dimensional scalar field. As the structure of the action makes it clear that it cannot be decoupled using a linear transformation, we will work with the coupled system as a whole.

\subsection{Quantum mechanical analogy} 

The first observation we can make about the system~\eqref{system} is that in the zero-gravity limit~($\beta=0$) it decouples and simplifies greatly:
$$
\begin{cases}
\displaystyle \hfill-h''=\tilde\mu^2 h \ ,\cr
\displaystyle -f ''+ \frac{\Phi'''}{\Phi'}f =\mu^2 f \ ,
\end{cases}
$$
which allows us to find its solutions immediately. The solutions of the upper equation are simply the plane waves; the second equation, determining~$f$, reduces to Eqn.~\eqref{kink_wave_f} and therefore its solutions are the familiar normal modes of the kink, including the normalized zero mode.

In general, when $\beta\neq0$, the problem of finding the solutions of~\eqref{system} is far less trivial. Eqn.~\eqref{system} being a system of coupled second order differential equations with fairly complicated coefficients, solving it exactly is certainly not an option. It is obvious that eliminating one of the functions, passing to a fourth order equation is possible, but by no means would it simplify our task. It is preferable to consider the problem of finding the solutions of~\eqref{system} as the eigenvalue problem
\begin{equation}
\label{eigenproblem}
{\cal H}_1 \Psi_\mu=\mu^2 \Psi_\mu \ ,\qquad \Psi_\mu\equiv\begin{pmatrix}h_\mu \cr f_\mu\end{pmatrix}\ ,
\end{equation}
for the matrix Schr\"odinger operator: 
$$
{\cal H}_1=e^A
\begin{pmatrix}\displaystyle -\frac{d^2}{dr^2}-2A'\frac{d}{dr}
-\left(2A''+A'^2\right)& \displaystyle\sqrt{\frac{2\kappa}{3}}\left(2\Phi''+\Phi'A'\right) 
\\[1.5ex]
 \displaystyle\sqrt{\frac{2\kappa}{3}}\left(2\Phi''+\Phi'A'\right) & \displaystyle \quad-\frac{d^2}{dr^2}-2A'\frac{d}{dr}+\left(\frac{\Phi'''}{\Phi'}+2A'\frac{\Phi''}{\Phi'}-A''\right)\end{pmatrix} \ .
$$
(In the above expressions the index $\mu$ denotes the eigenvalue and should not be confused for a vector index.) 
For any given~$\mu$, there are four linearly independent solutions of~\eqref{system}. The form of~${\cal H}_1$ implies that its eigenfunctions are normalized with weight $e^{A(r)}$ (the precise form of the orthogonality relation will be given below). 

The quantum mechanical analogy suggests that we use methods inspired by the supersymmetric quantum mechanics~\cite{susy_qm}. As we will see, this approach greatly simplifies the problem of investigating the spectrum of~${\cal H}_1$. To begin, it can be easily checked that~${\cal H}_1$ can be written in a factorized form as~${\cal H}_1={\cal A}^\dagger{\cal A}$, with:
\begin{align*}
{\cal A}&=e^{-{A}/{4}} \begin{pmatrix}\displaystyle \frac{d}{dr}+\frac{1}{4}A' &\qquad\displaystyle -\sqrt{\frac{2\kappa}{3}}\Phi' \\[1.5ex] \displaystyle -\sqrt{\frac{2\kappa}{3}}\Phi' & \qquad\displaystyle\frac{d}{dr}-\frac{3}{4}A'-\frac{\Phi''}{\Phi'}\smallskip\end{pmatrix} e^{{3}A/4}\ , \\[1ex]
{\cal A}^\dagger&=e^{-{A}/{4}} \begin{pmatrix}\displaystyle -\frac{d}{dr}+\frac{1}{4}A' & \quad\displaystyle-\sqrt{\frac{2\kappa}{3}}\Phi' \\[1.5ex] \displaystyle-\sqrt{\frac{2\kappa}{3}}\Phi' & \quad\displaystyle-\frac{d}{dr}-\frac{3}{4}A'-\frac{\Phi''}{\Phi'}\smallskip\end{pmatrix} e^{{3}A/4} \ , 
\end{align*}
where the exponential factors are related with the choice of coordinates and a non-trivial weight function present in~\eqref{system}. 

\subsection{Nature of the spectrum} 

Let us investigate whether the spectrum of the operator~${\cal H}_1$ contains any bound states. 
 The factorization~${\cal H}_1={\cal A}^\dagger{\cal A}$ guarantees that the eigenvalues of~${\cal H}_1$ are non-negative and allows one to find easily the~$\mu=0$ solutions of the system~\eqref{system}. Let us therefore start by checking for presence of a localized zero mode. 
The zero mode of~${\cal H}_1$, if there is one, should satisfy either:
\begin{equation}
\label{zero_hom}
{\cal A}\begin{pmatrix}h_0 \cr f_0\end{pmatrix}=0 \ .
\end{equation}
or an inhomogenous equation 
\begin{equation}
\label{zero_inhom}
{\cal A}\begin{pmatrix}h_0 \cr f_0\end{pmatrix}=\begin{pmatrix}\hhh_0 \cr \fff_0\end{pmatrix} \ ,
\end{equation}
where the inhomogenous term is one of the solutions 
of ${\cal A}^\dagger\Bigl(\begin{smallmatrix}\hhh_0 \cr \fff_0\end{smallmatrix}\Bigr)=0$~(see Appendix~\ref{app-mu=0_sol}). 
The general solution of both the above equations can be calculated exactly. The homogeneous solution reads:
\begin{equation}
\label{Psi_0}
\Psi_0(r)=\begin{pmatrix}h_0 \cr f_0\end{pmatrix}
=C_1^{(0)}\begin{pmatrix}A'e^{-A} \\[1.5ex] \displaystyle-\sqrt{\frac{2\kappa}{3}}\Phi'e^{-A}\smallskip\end{pmatrix} + C_2^{(0)}\begin{pmatrix}\displaystyle 1-A'e^{-A}\int_0^r dy e^{A(y)} \\[1.5ex] \displaystyle\sqrt{\frac{2\kappa}{3}}\Phi'e^{-A}\int_0^r dy e^{A(y)}\smallskip\end{pmatrix} \ .
\end{equation}
Let us remark that the solution~$C_1^{(0)}$ can be derived as a coordinate transformation:
$$
 \delta_\xi h_{MN}=-\xi_{N;M}-\xi_{M;N}
$$
with~$\xi_\mu=0$ and~$\xi_r=e^{-A}$. 
 For our background, the solution~\eqref{Psi_0} is clearly not normalizable; for any choice of~$C_1^{(0)}$ and~$C_2^{(0)}$ we have:
$$
\int_{-\infty}^\infty dr\, e^{A(r)} \Psi_0^T(r)\Psi_0^{}(r)= \int_{-\infty}^\infty dr\, e^{A(r)} \left[h_0^2(r)+f_0^2(r)\right] =\infty \ .
$$
This remains true for the two remaining~$\mu=0$ solutions of~\eqref{system}, which can be determined from~\eqref{zero_inhom} (as the explicit expressions for these solutions are fairly complicated and not very illuminating, we relegate them to Appendix~\ref{app-mu=0_sol}, where the determination of the solutions of~${\cal H}_1 \Psi_0=0$ is presented in more detail). There is therefore no localized zero mode in our model.

Let us now turn to the~$\mu\neq0$ spectrum. To determine the nature of this spectrum, it is enough to determine the asymptotic behavior of solutions. Supposing~$\beta\ll 1$, far from the core of the domain wall, for positive $r$'s, the system~\eqref{system} becomes:
\begin{equation}
\begin{cases}
\displaystyle \hfill-h'' +8\xi^2 h'-16\xi^4 h -4\sqrt{6\beta}\left(a^2 +a\xi^2\right) e^{-2ar}f=\mu^2 e^{4\xi^2r} h\ , \cr
\displaystyle -f '' +8\xi^2 f'+\left(4a^2+16a\xi^2\right)f-4\sqrt{6\beta}\left(a^2 +a\xi^2\right) e^{-2ar}h =\mu^2 e^{4\xi^2r}f \ ,
\end{cases}
\end{equation}
where~$\xi^2\equiv \beta a$. For~$|r|\gg1/a$ the mixing term is clearly negligible and we are left with a simple system of decoupled equations, whose solutions can be written in terms of the Bessel functions:
\begin{align*}
\Psi_\mu\mathop{\longrightarrow}_{r\to\infty}
&\begin{pmatrix}e^{4\xi^2r}\left[D_1J_0\!\left(\frac{\mu}{2\xi^2}e^{2\xi^2r}\right)+D_2Y_0\!\left(\frac{\mu}{2\xi^2}e^{2\xi^2r}\right)\right]\smallskip\cr 0\end{pmatrix}\\
& +\begin{pmatrix}0\smallskip\cr e^{4\xi^2r}\left[ D_3J_{\nu+1}\!\left(\frac{\mu}{2\xi^2}e^{2\xi^2r}\right)+D_4Y_{\nu+1}\!\left(\frac{\mu}{2\xi^2}e^{2\xi^2r}\right)\right]\end{pmatrix} \ ,
\end{align*}
where~$\nu=1/\beta+1$. The continuous spectrum of~${\cal H}_1$ starts therefore right above zero. There are no localized modes. 

\subsection{Partner system and the determination of the eigenfunctions}

Absence of bound states in the spectrum of~${\cal H}_1$ for~$\beta\neq0$ does not necessarily mean that the localized zero mode present in the non-gravitating case~$\beta=0$ has altogether disappeared as soon as the gravity was switched on. It is very likely that in the presence of gravity the would-be zero mode has become a quasilocalized state. To investigate this possibility, it is necessary to know the eigenfunctions of~${\cal H}_1$ for non-zero~$\mu$ in some more detail. 
To this purpose, it is very convenient to define the ``partner'' operator ${\cal{H}}_2={\cal A}{\cal A}^\dagger$. This operator is much easier to study, as its eigenvalue problem
$${\cal{H}}_2\begin{pmatrix}\hh_\mu \cr \ff_\mu\end{pmatrix}=\mu^2\begin{pmatrix}\hh_\mu \cr \ff_\mu\end{pmatrix}$$
has the advantage of being a system of two \emph{decoupled} equations:
\begin{equation}
\label{partner}
\begin{cases}
\displaystyle \hfill-\hh'' - 2A'\hh'-\left(\frac{3}{2}A''+\frac{3}{4}A'^2\right)\hh =\mu^2 e^{-A}\hh\ , \\[1.5ex]
\displaystyle-\ff '' - 2A'\ff'-\left[\frac{\Phi'''}{\Phi'}-2\left(\frac{\Phi''}{\Phi'}\right)^2-A'\frac{\Phi''}{\Phi'}+\frac{5}{2}A''+\frac{3}{4}A'^2\right]\ff =\mu^2 e^{-A}\ff\ .
\end{cases}
\end{equation}
At the same time, once we know the eigenfunctions of the partner operator~${\cal H}_2$, we automatically know the eigenfunctions of~${\cal H}_1$. Indeed, there is a one-to-one correspondence between the eigenfunctions of the two operators: for any~$\mu\neq0$, the solutions of the coupled system can be determined from the solutions of the partner system using:
\begin{equation}
\label{recipe}
\Psi_\mu\equiv \begin{pmatrix}h_\mu \cr f_\mu\end{pmatrix}=\frac{1}{\mu}\, {\cal A}^{\dagger}\!\begin{pmatrix}\hh_\mu \cr \ff_\mu\end{pmatrix}=\frac{1}{\mu}\,\begin{pmatrix}e^{A/2}\left[-\hh'_\mu-\frac12A'\hh_\mu\right] -\sqrt{\frac{2\kappa}{3}}e^{A/2}\Phi'\ff_\mu \\[1.5ex] e^{A/2}\left[-\ff_\mu'-\left(\frac32A'+\frac{\Phi''}{\Phi'}\right)\ff_\mu\right]-\sqrt{\frac{2\kappa}{3}}e^{A/2}\Phi'\hh_\mu\end{pmatrix} \ .
\end{equation}
Notice that if $\tilde\Psi_\mu\equiv\Bigl(\begin{smallmatrix}\hh_\mu \cr \ff_\mu \end{smallmatrix}\Bigr)$ is normalized, so is $\Psi_\mu$ (see the Appendix~\ref{app-connection} for more details). We will use the notation:
\begin{equation}
\label{basis} 
\Psi_{\mu(\hh)}=\frac{1}{\mu}\,{\cal A}^{\dagger}\!\begin{pmatrix}\hh_\mu \cr 0\end{pmatrix}\ , \qquad \qquad
 \Psi_{\mu(\ff)}=\frac{1}{\mu}\, {\cal A}^{\dagger}\!\begin{pmatrix}0 \cr \ff_\mu\end{pmatrix} \ ,
\end{equation}
where the eigenfunctions~$\Psi_{\mu(\alpha)}$, with~$\alpha\in\{\hh,\ff\}$, satisfy the following orthogonality relation:
\begin{equation}\label{ortho_psi}
\int_0^\infty dr \, e^{A(r)} \,\Psi_{\mu(\alpha)}^T(r)\Psi^{}_{\mu'(\beta)}(r)=\delta_{\alpha\beta}\delta(\mu-\mu')\ .
\end{equation}

\FIGURE[ht]{\includegraphics[width=\textwidth]{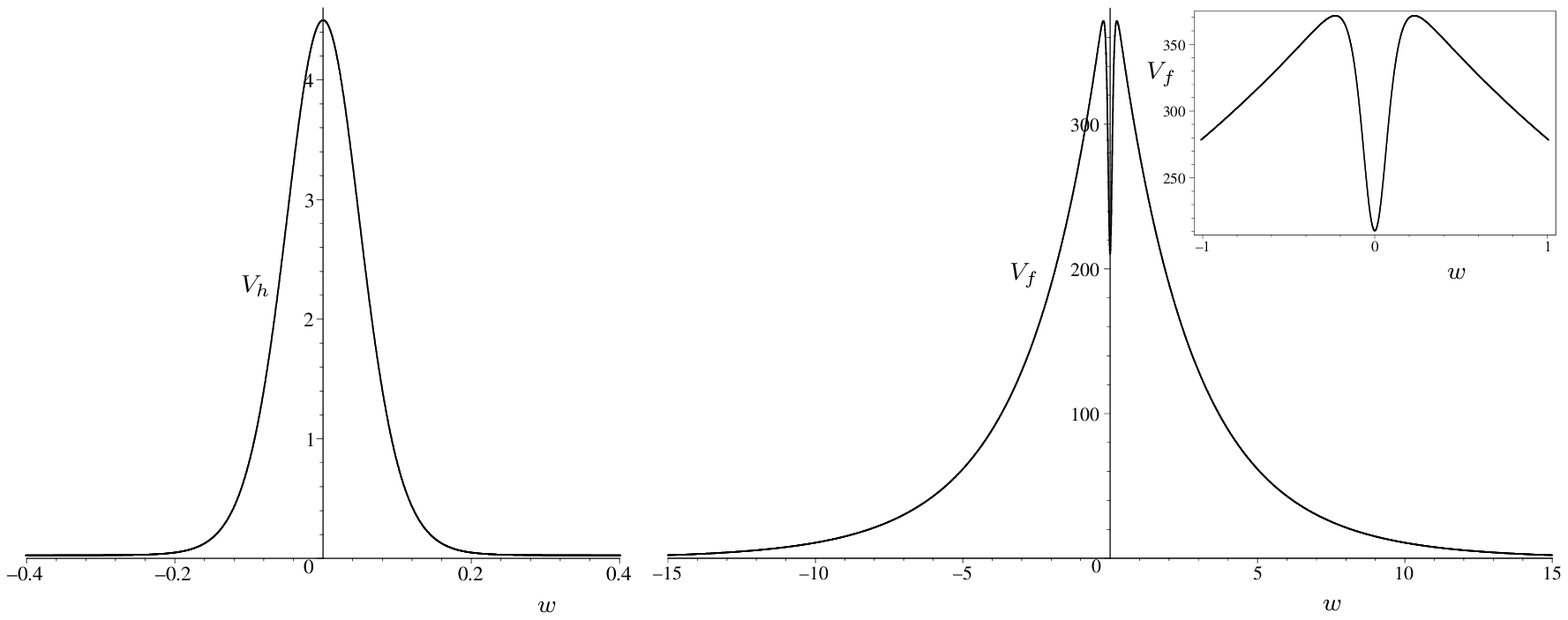}%
\caption{{\label{fig-potentials}}Behavior of the Schr\"odinger potentials of the partner system: to the left potential for~$\hh$, to the right potential for~$\ff$. The plots are made in the conformal coordinates, $w=\int^r dr' e^{-A(r')/2}$. The parameter values are~$\beta=10^{-2},\, a=10$.}}

Let us therefore turn to the spectrum of the partner operator~${\cal{H}}_2$. Already from the pictures of the Schr\"odinger potentials determining the wave functions~(see Fig.~\ref{fig-potentials}) it transpires that the spectrum of~${\cal{H}}_2$ is, as expected, purely continuous. As for the eigenfunctions, even if the partner system~\eqref{partner} is considerably reduced in complexity compared to~\eqref{system}, they cannot be calculated exactly for non-zero~$\mu$. It is however possible to find the solutions of~\eqref{partner} in the thin wall limit~\eqref{tw}. The determination of solutions being somewhat lengthy, we give here only its main points and the results relevant for the forthcoming considerations, refering the interested readers to the Appendix~\ref{app-eigenfunctions} for the details of the calculations.

The solution becomes particularly elegant and straightforward for the upper of the equations~\eqref{partner}, determining the evolution of~$\hh$. Indeed, in the thin wall limit this equation becomes:
\begin{equation}
-\hh'' + 8 \xi^2\mathop{\rm sign}(r)\hh'+\left(12\xi^2\delta(r)- 12\xi^4\right)\hh ={\mu}^2 e^{4\xi^2|r|}\hh 
\end{equation} 
and its general solution for arbitrary~$\mu$ can be found. It can always be written as a linear combination of an even and an odd solution. It is the even, appropriately normalized,~$\hh_\mu$ solution which will be of interest for what follows. It is of the following form:
\begin{equation}
\label{hh_even_explicit}
\hh_\mu^{(e)}(r)= A^{(e)}e^{4\xi^2 |r|}\left[Y_2\!\left(\frac{\mu}{2\xi^2}\right)J_1\!\left(\frac{\mu}{2\xi^2}e^{2\xi^2 |r|}\right)-J_2\!\left(\frac{\mu}{2\xi^2}\right) Y_1\!\left(\frac{\mu}{2\xi^2}e^{2\xi^2 |r|}\right)\right] \ ,
\end{equation} 
with the normalization constant~$A^{(e)}$ given by: 
\begin{equation}
\label{A_e}
A^{(e)}=\frac{\sqrt{\mu}}{2\xi}\left[J_2^2\!\left(\frac{\mu}{2\xi^2}\right)+Y_2^2\!\left(\frac{\mu}{2\xi^2}\right)\right]^{-1/2}\ .
\end{equation}
The second equation, which determines the evolution of~$\ff_\mu$, becomes for $|r|\gg1/a$:
\begin{equation}
-\ff '' +8\xi^2\mathop{\rm sign}(r)\ff'+\left(4a^2+8\xi^2a-12\xi^4\right)\ff=\mu^2 e^{4\xi^2 |r|}\ff \ .
\end{equation} 
Again, we can easily find the solutions of this equation, however they cannot be extended towards~$r=0$ as simply as the~$\hh_\mu$ solutions. The (approximate) global solutions of this equation can be found for small~$\mu$ through matching of the~$|r|\gg1/a$ solutions obtained in the thin wall limit to the exact~$\mu=0$ solutions. In what follows, we will need the odd~$\ff_\mu$ (again, properly normalized):
\begin{equation}
\label{ff_odd_explicit}
\ff^{(o)}(r)\approx \frac{\sqrt{\mu}}{2\xi}
\begin{cases}\displaystyle -\frac{v}{2\pi}\left(\frac{e\mu}{4\nu\xi^2}\right)^{\nu}\,e^{-3A/2}\frac{A'}{\Phi'} & |r|<d \ , \\[1.5ex] \displaystyle 
\,\mathop{\rm sign}(r)\, e^{4\xi^2 |r|}J_\nu\left(\frac{\mu}{2\xi^2}e^{2\xi^2 |r|}\right) & |r|\gg\displaystyle\frac{1}{a}\ ,
\end{cases}
\end{equation} 
where~$d\sim{\cal{O}}(\xi^{-2})$. The latter solution is valid for~$\mu\lesssim \nu \xi^2\approx a$. 

\section{\label{sect-static}Static potential between sources on the brane}

In the presence of other matter fields, the system of the gravitational and scalar perturbations couples to the sources via the interaction term:
\begin{equation}
\label{S_source}
{\cal{S}}_{\mathrm{int}}=\int d^4x\, dr \sqrt{g}\left[\frac{1}{2}h^{MN}T_{MN}+J\phi\right]\ .
\end{equation}
Supposing that the coupling of these matter fields to the gravity is much weaker than to the scalar field and its perturbations, we restrict our attention to the sources of the field~$\phi$. This means that our scalar system~\eqref{system} is coupled to the sources of the form~$\underline{J}=\bigl(\begin{smallmatrix}0\cr J_f\end{smallmatrix}\bigr)$:
\begin{equation}
\left\{\begin{aligned}
\displaystyle e^{-A}\eta^{\mu\nu}\partial_\mu \partial_\nu \chi -\chi'' - 2A'\chi'-(2A''+A'^2)\chi +\sqrt{\frac{2\kappa}{3}}\!\left(2\Phi''+\Phi'A'\right)\phi&=0\ , \cr
\displaystyle \! e^{-A}\eta^{\mu\nu}\partial_\mu \partial_\nu \phi-\phi '' - 2A'\phi'+\left[\frac{\Phi'''}{\Phi'}+2A'\frac{\Phi''}{\Phi'}-A''\right]\!\phi+\sqrt{\frac{2\kappa}{3}}\!\left(2\Phi''+\Phi'A'\right)\chi &=J_f(x,r) \ .
\end{aligned}\right.\!\!\!\!\!
\label{system_chi_phi}
\end{equation}
In this section, we will study the behavior of the potential between two static sources $\underline{J}_{(1)}$ and $\underline{J}_{(2)}$ of the field~$\phi$.

We suppose that the sources are pointlike and situated at $r=r'=0$:
\begin{align*}
\underline{J}_{(1)}(\vec{x},r)&=\underline{\tilde{J}}_{(1)}\,\delta^3(\vec{x}-\vec{x}_1)\delta(r)\ , \\
\underline{J}_{(2)}(\vec{x},r)&=\underline{\tilde{J}}_{(2)}\,\delta^3(\vec{x}-\vec{x}_2)\delta(r) \ .
\end{align*} 
The static potential between $\underline{J}_{(1)}$ and $\underline{J}_{(2)}$ receives Yukawa-type contributions from all modes and is given by the following integral (see Appendix~\ref{app-static} for the calculations leading to the expression below):
\begin{align}
V(|\vec{x}_1-\vec{x}_2|)&=\frac{1}{4\pi}\int_0^\infty d\mu \frac{e^{-\mu|\vec{x}_1-\vec{x}_2|}}{|\vec{x}_1-\vec{x}_2|}\sum_\alpha\underline{\tilde{J}}_{(1)}^T\Psi_{\mu(\alpha)}(0)\Psi^T_{\mu(\alpha)}(0)\underline{\tilde{J}}_{(2)} \nonumber \\
&=\frac{1}{4\pi}{\tilde{J}}_{(1),f}{\tilde{J}}_{(2),f}\int_0^\infty d\mu \frac{e^{-\mu|\vec{x}_1-\vec{x}_2|}}{|\vec{x}_1-\vec{x}_2|} \sum_\alpha \left[f_{\mu(\alpha)}(0)\right]^2 \ ,
\label{potential}
\end{align}
where $\Psi_{\mu(\alpha)}$, with $\alpha\in\{\hh,\ff\}$ are the eigenfunctions~of the operator~${\cal H}_1$ satisfying the normalization condition~\eqref{ortho_psi}
and the following boundary conditions at~$r=0$:
\begin{align}
\label{bc}
[\Psi_{\mu(\alpha)}]_1(0)&=h_{\mu(\alpha)}(0)=0\ , \nonumber \\
\partial_r[\Psi_{\mu(\alpha)}]_2(0)&= \partial_r f_{\mu(\alpha)}(0)=0 \ .
\end{align} 
These are simply
$$
\Psi_{\mu(\hh)}=\frac{1}{\mu} {\cal A}^{\dagger}\begin{pmatrix}\hh_\mu^{(e)} \cr 0\end{pmatrix} \qquad \mbox{and}\qquad
 \Psi_{\mu(\ff)}=\frac{1}{\mu} {\cal A}^{\dagger}\begin{pmatrix}0 \cr \ff_\mu^{(o)}\end{pmatrix} \ ,
$$
with~$\hh^{(e)}_\mu$ given by~\eqref{hh_even_explicit} and~$\ff_\mu^{(o)}$ given by~\eqref{ff_odd_explicit}. For these we have (see Appendix~\ref{app-sol-coupled}):
\begin{align}
f_{\mu(\hh)}(0)&\approx \left(\frac{3}{\pi^2}\frac{1}{\beta}\right)^{1/2}\left(\frac{2\xi^2}{\mu}\right)^{3/2}\left[J_2^2\!\left(\frac{\mu}{2\xi^2}\right)+Y_2^2\!\left(\frac{\mu}{2\xi^2}\right)\right]^{-1/2}\ , \label{f-h}\\
f_{\mu(\ff)}(0)&\approx \left(\frac{9\nu}{16e}\right)^{1/2}\left(\frac{\mu e}{4\xi^2\nu}\right)^{\nu-1/2}\approx \left(\frac{9\nu}{16e}\right)^{1/2}\left(\frac{\mu e}{4a}\right)^{1/\beta} \ .\label{f-f}
\end{align}
The first expression is valid for an arbitrary $\mu$, the second one for $\mu\lesssim \nu \xi^2\approx a$. In all this interval~$f_{\mu(\ff)}$ is very strongly suppressed and its contribution to the potential can be safely neglected compared to the contribution of~$f_{\mu(\hh)}$. Considering distances~$R\gg1/a$, we can keep infinity as the upper limit of integration. The potential therefore becomes:
\begin{align}
V(R) &\approx \frac{1}{4\pi}\tilde{J}_{(1),f}\tilde{J}_{(2),f}\int_0^\infty d\mu \,\frac{e^{-\mu R}}{R}\frac{3}{\pi^2}\frac{1}{\beta}\left(\frac{2\xi^2}{\mu}\right)^3\left[J_2^2\!\left(\frac{\mu}{2\xi^2}\right)+Y_2^2\!\left(\frac{\mu}{2\xi^2}\right)\right]^{-1}\nonumber \\
&= \tilde{J}_{(1),f}\tilde{J}_{(2),f}\ \frac{3a}{2\pi^3 R}\int_0^\infty d\alpha \,\frac{e^{-(2\xi^2 R) \alpha}}{\alpha^3 \left[J^2_2(\alpha) +Y^2_2(\alpha)\right]} \ ,
\label{potential_explicit}
\end{align}
where $R\equiv|\vec{x}_1-\vec{x}_2|$ and where we have introduced a dimensionless variable~$\alpha=\mu/(2\xi^2)$. Remarkably, apart from the factor in front of the integral (identical to the non-gravitating case) the behavior of the static potential~\eqref{potential_explicit} does not depend on the details of the brane structure, but solely on the geometry of the space-time. Indeed, the integral in~\eqref{potential_explicit} contains only one parameter,~$2\xi^2 R$. This means that it is the~$AdS$ radius of the five-dimensional space-time,~$(2\xi)^{-2}$, which will determine how the potential~$V(R)$ changes with the distance between sources.

Before actually estimating~$V(R)$, it is worthwhile to take a closer look at the spectral density function~$\rho(\alpha)=\alpha^{-3} \left[J^2_2(\alpha) +Y^2_2(\alpha)\right]^{-1}$, which should reveal whether there are any resonances present in our system. The behavior of this function is presented on Fig.~\ref{fig-density}. The bulk of the weight of~$\rho(\alpha)$ is concentrated in the region of fairly light modes, $\mu\sim {\cal{O}}(\xi^2)$ (or $\alpha\sim{\cal{O}}(1)$), the width of the peak being of the same order as its distance from the origin. Thus, from the point of view of the four-dimensional physics the collection of modes with masses~$\mu\sim {\cal{O}}(\xi^2)$ behaves as a wide resonance. This phenomenon is unlike any others previously observed in the context of brane scenarios. Indeed, let us remind that in the case of quasilocalized massive scalars considered in~\cite{drt} the spectral density function exhibits characteristic features of a narrow resonance: a sharp peak around~$\mu=m/\sqrt{2}$,~$m$ being the mass of the bulk scalar; the width of the peak is suppressed by a factor~$(m/\xi^2)^2$. The signature of a quasilocalized massless particle would be a density function steeply concentrated in the vicinity of the origin~\cite{ceh,us-toy}. 

\FIGURE[h]{\includegraphics[width=0.6\textwidth]{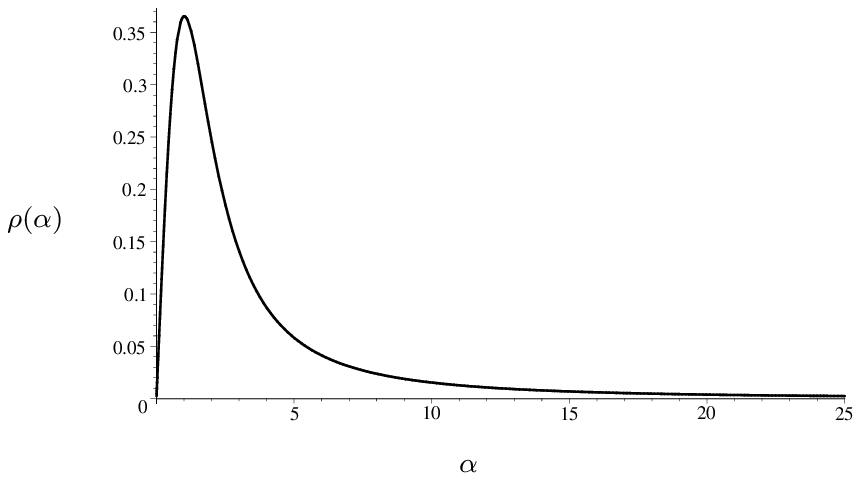}%
\caption{{\label{fig-density}}Behavior of the function~$\rho(\alpha)=\alpha^{-3} \left[J^2_2(\alpha) +Y^2_2(\alpha)\right]^{-1}$.}}

Let us now proceed to determine how the potential~\eqref{potential_explicit} depends on the distance between sources (to simplify the notation, from now on we will set $\tilde{J}_{(1),f}=\tilde{J}_{(2),f}=1$). The dominating behavior of~$V(R)$ can be determined at the opposite ends of distance scale, for~$2\xi^2R\gg1$ and for~$\xi^2/a\ll2\xi^2R\ll1$. At large distances, $2\xi^2R\gg1$, only small $\alpha$'s will contribute, and we can approximate the integrand using the expansion of the Bessel functions for small argument. The potential $V(R)$ becomes then:
\begin{align}
V(R) &\approx \frac{3a}{16\pi R}\int_0^\infty d\alpha \,e^{-(2\xi^2 R)\alpha} \, \frac{\alpha}{2+\alpha^2} \nonumber \\
& = \frac{3a}{16\pi R}\left[-\mathop{\rm ci}(2\sqrt{2}\xi^2R)\cos(2\sqrt{2}\xi^2R) - \mathop{\rm si}(2\sqrt{2}\xi^2R)\sin(2\sqrt{2}\xi^2R)\right]\nonumber \\
 &\approx \frac{3a}{16\pi R}\left[\frac{1}{2}\frac{1}{(2\xi^2R)^2} -\frac{3}{2}\frac{1}{(2\xi^2R)^4}+\dots\right]\ ,
\label{pot_far}
\end{align}
where the last line was obtained by asymptotic expansion, using $2\xi^2R\gg 1$. Hence, in this region of distances the static potential between two sources has a~$1/R^3$ fall-off and its amplitude is strongly suppressed. The approximate expression for~\eqref{pot_far} is plotted on Fig.~\ref{fig-pot-R} together with the numerically evaluated~$V(R)$, demonstrating that the analytical result is an excellent estimation of the potential's behavior at large distances. 

Let us now estimate the potential at small distances, $\xi^2/a\ll2\xi^2R\ll1$. We can rewrite~\eqref{potential_explicit} in the following way:
\begin{align}
V(R) \approx \frac{3a}{2\pi^3 R}\Biggl\{&\int_0^{\infty}d\alpha\,\frac{1}{\alpha^3 \left[J^2_2(\alpha) +Y^2_2(\alpha)\right]}+\int_0^{\infty}d\alpha\,\frac{e^{-2\xi^2 R\alpha}-1}{\alpha^3 \left[J^2_2(\alpha) +Y^2_2(\alpha)\right]}\Biggr\} \nonumber \\
\approx \frac{3a}{2\pi^3 R}\Biggl\{&\int_0^{\infty}d\alpha\,\frac{1}{\alpha^3 \left[J^2_2(\alpha) +Y^2_2(\alpha)\right]}-\frac{\pi}{2}\int_0^\infty d\alpha \,\left(1-e^{-2\xi^2 R\alpha}\right)\frac{1}{1+\alpha^2} \nonumber \\
&+\int_0^\infty d\alpha\left[\frac{e^{-2\xi^2 R\alpha}-1}{\alpha^2 \left[J^2_2(\alpha) + Y^2_2(\alpha)\right]}-\frac{\pi}{2}\frac{1}{\alpha^2+1}\left(e^{-2\xi^2 R\alpha}-1\right)\right]\Biggr\} \nonumber \\
\approx \frac{3a}{2\pi^3 R}\Biggl\{&\int_0^{\infty}d\alpha\,\frac{1}{\alpha^3 \left[J^2_2(\alpha) +Y^2_2(\alpha)\right]}-\frac{\pi}{2}\int_0^\infty d\alpha \,\left(1-e^{-2\xi^2 R\alpha}\right)\frac{1}{1+\alpha^2} \nonumber \\
&- 2\xi^2 R \int_0^\infty d\alpha\left[\frac{1}{\alpha^2 \left[J^2_2(\alpha) + Y^2_2(\alpha)\right]}-\frac{\pi}{2}\frac{\alpha}{\alpha^2+1}\right]\Biggr\}\ . 
\label{V_expansion}
\end{align}
While the first integral in~\eqref{V_expansion} does not seem to be tabulated, its numerical evaluation indicates very suggestively that:
\begin{equation}
\label{exact_int}
\int_0^{\infty}d\alpha\,\frac{1}{\alpha^3 \left[J^2_2(\alpha) +Y^2_2(\alpha)\right]}=\frac{\pi^2}{8} \ .
\end{equation}
The second integral can be expressed through the sine and the cosine integrals:
\begin{align*}
\frac{\pi}{2}\int_0^\infty d\alpha \,\left(e^{-2\xi^2 R\alpha}-1\right)\frac{1}{1+\alpha^2} &=\frac{\pi}{2}\left[\mathop{\rm ci}(2\xi^2R)\sin(2\xi^2R)-\mathop{\rm si}(2\xi^2R)\cos(2\xi^2R)-\frac{\pi}{2}\right] \\
&\approx \pi\xi^2R\ln(\xi^2R)+\pi\xi^2R\left(\ln(2)+\gamma-1\right)\ ,
\end{align*}
where the second line was obtained using the series expansion. Inserting these results into~\eqref{V_expansion} we obtain
\begin{equation}
V(R) \approx \frac{3a}{16\pi R}\left[1+\frac{8}{\pi}\xi^2R\ln(\xi^2R)+\frac{8}{\pi}\xi^2R\left(\gamma-1+\ln(2)-\frac{2}{\pi}C_1\right)\right] \ ,
\label{potential_cont}
\end{equation}
where the constant $C_1$ stands for the result of the third integral, which must be found numerically: $C_1\approx-0.804$. The comparison between this expression and the numerically evaluated~$V(R)$ is presented on~Fig.~\ref{fig-pot-R}. The way in which the above expression depends on the strength of gravity is very natural: as~$\xi^2\equiv\beta a$, for a fixed~$a$, the smaller is the gravitational constant, the larger the interval~$\xi^2/a\ll2\xi^2R\ll1$ and the closer the above expression for the potential~$V(R)$ becomes to~${3a}/(16\pi R)$, which is the form of the static potential in the zero-gravity ($\beta=0$) case. When the gravity is switched off, the structure of the spectrum changes radically: the fields $\chi$ and $\phi$ get completely decoupled and $\phi$ has a normalizable zero mode:
$$
f_0(r)=\frac{\sqrt{3a}}{2} \frac{1}{\cosh^2(ar)}u_0(x) \ .
$$
The potential mediated by the field~$u_0$ between two static pointlike sources on the brane is in this case:
$$
V(R)=\frac{1}{4\pi R}\left[f_0(0)\right]^2 = \frac{3a}{16\pi R} \ .
$$
The potential~\eqref{potential_explicit} has therefore a continuous limit towards the non-gravitating case (see~Fig.~\ref{fig-pot-short}). To conclude, let us observe that the fact that we find the~$1/R$ term in the region of distances small compared to the~$AdS$ radius is not at all unexpected, as at these scales the space-time looks practically flat and the background solution~\eqref{Phi}--\eqref{U} is very close to the setup of~Ref.~\cite{shap}.

\FIGURE[t]{\includegraphics[width=\textwidth]{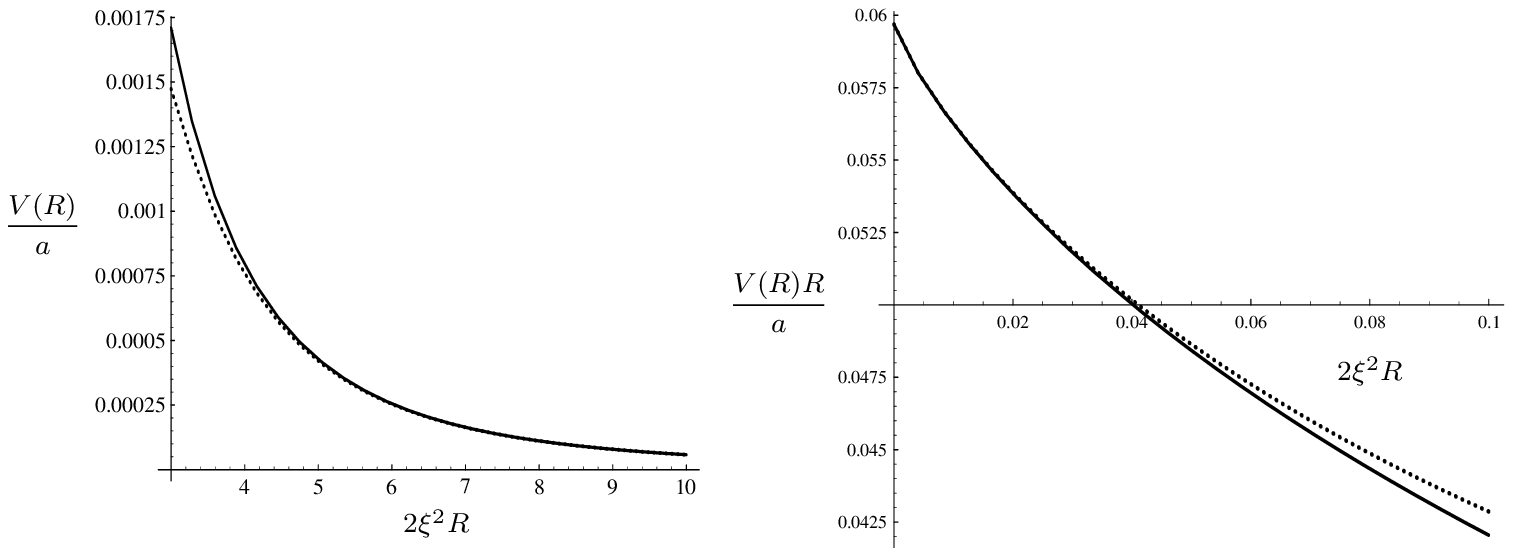}%
\caption{{\label{fig-pot-R}}Behavior of the potential as a function of~$R$: to the left for the large distances, to the right for the short distances. The continuous line is the numerical evaluation of the potential, the dotted line represents the analytical estimation.}}

\FIGURE[t]{\includegraphics[width=0.6\textwidth]{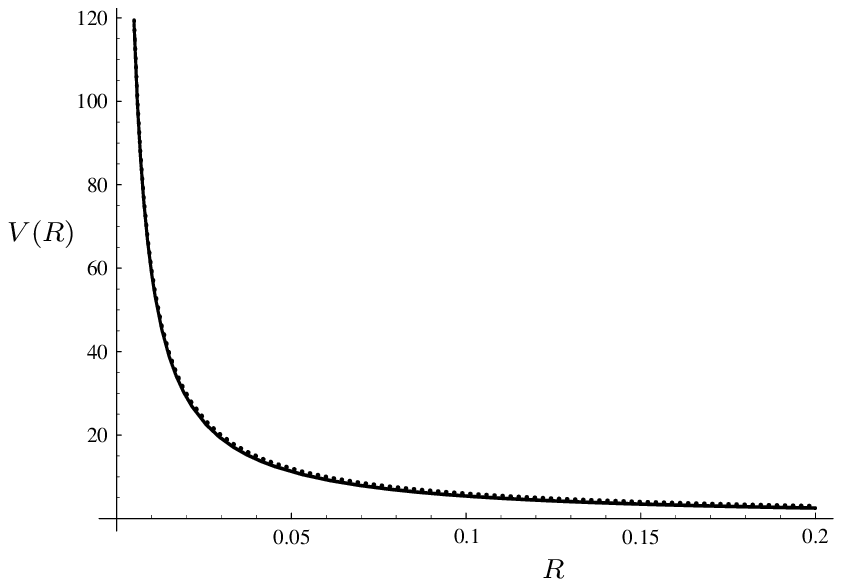}%
\caption{{\label{fig-pot-short}}Behavior of the potential at short distances: the dotted line is the~$\beta=0$ case, the continuous line represents the gravitating case. The parameter values are~$\beta=10^{-2},\, a=10$.}}

\section{\label{sect-periodic}Periodic source on the brane}

In the previous section we have considered static sources. However, in the actual fact~$\phi$'s are four-dimensional particles and the source term in~\eqref{S_source} is simply an effective description of their interactions with other fields localized on the brane. In order to get some insight into the dynamics of these particles, we must extend our study to time-dependent sources. 

In this section, we will study the propagation of signal emitted by an oscillating source of~$\phi$. Let us therefore consider a pointlike source of the form $\bigl(\begin{smallmatrix}0\cr J_f\end{smallmatrix}\bigr)$ situated on the brane (more specifically, we consider the hypersurface~$r=0$) and oscillating with frequency~$\omega$. 
 The field induced on the brane is given by the convolution of the source with the Green's function:
\begin{equation}
{\mathbf G}_\omega(\vec{x}-\vec{x}')=\int_{-\infty}^{\infty}d(t^\prime-t) \ {\mathbf G}(x,r=0;x',r'=0) \ e^{-i\omega (t-t^\prime)}\ ,
\end{equation}
where~${\mathbf G}(x,r;x',r')$ is the retarded Green's function of Eqn.~\eqref{system_chi_phi}, given by
\begin{equation}
{\mathbf G}(x,r;x',r')=\int \frac{d^4k}{(2\pi)^4}e^{ik_\mu(x^\mu -x'^\mu)}\int_0^\infty d\mu\, \sum_\alpha \,\frac{\Psi_{\mu(\alpha)}(r)\Psi_{\mu(\alpha)}^T(r')}{\mu^2-k^2-2i\varepsilon k_0}\ ,
\end{equation}
where, as in the previous section~$\Psi_{\mu(\alpha)}$ are the eigenfunctions of the operator~${\cal{H}}_1$, satisfying the boundary conditions~\eqref{bc}. As $h_\mu(r)$ components are odd, purely $f$-type field will be produced; it is determined by the $(f,f)$-component of the Green's function. After inserting~${\mathbf G}_{(f,f)}(x,r;x',r')$ and simplifying we obtain
\begin{equation}
\label{G_omega}
{\mathbf G}^\omega_{(f,f)}(R)\equiv{\mathbf G}^\omega_{(f,f)}(x,r=0;x',r'=0)=\frac{1}{4\pi}\int_0^\infty d\mu\ \frac{e^{i\omega_\mu R}}{R}\ \sum_\alpha\left[f_{\mu(\alpha)}(0)\right]^2 \ ,
\end{equation}
where $R\equiv|\vec{x}-\vec{x}^\prime|$ and where~$\omega_\mu=\sqrt{\omega^2-\mu^2}$ when~$\mu<\omega$ and~$\omega_\mu=i\sqrt{\mu^2-\omega^2}$ when~\mbox{$\mu>\omega$}. Only modes with~$\mu<\omega$ are actually radiated; the other ones exponentially fall off from the source. Being interested in the propagation of waves, we can integrate the expression for~${\mathbf G}^\omega_{(f,f)}(R)$ only up to~$\mu=\omega$. Limiting our interest to frequencies~$\omega\ll a$, we can neglect the contribution of~$f_{\mu(\ff)}$, given by~\eqref{f-f}, compared to the one of~$f_{\mu(\hh)}$, given by~\eqref{f-h}. ${\mathbf G}^\omega_{(f,f)}(R)$ becomes then:
\begin{align*}
{\mathbf G}_{(f,f)}^\omega(R)&\approx \frac{1}{4\pi} \int_0^\omega d\mu \,\frac{e^{i\omega_\mu R}}{R}\frac{3}{\pi^2}\frac{1}{\beta}\left(\frac{2\xi^2}{\mu}\right)^3\left[J_2^2\!\left(\frac{\mu}{2\xi^2}\right)+Y_2^2\!\left(\frac{\mu}{2\xi^2}\right)\right]^{-1} \\
&= \frac{3a}{2\pi^3 R}\int_0^{\omega/2\xi^2} \frac{d\alpha}{\alpha^3} \,\frac{\exp\left[i\omega R\sqrt{1-(2\xi^2/\omega)^2\alpha^2}\right]}{J^2_2(\alpha) +Y^2_2(\alpha)} \\
&= \frac{3a}{2\pi^3 R}\int_0^{\omega/2\xi^2} {d\alpha}\,{\rho(\alpha)}\exp\left[i\omega R\sqrt{1-(2\xi^2/\omega)^2\alpha^2}\right] \ .
\end{align*}
The above expression for~${\mathbf G}^\omega_{(f,f)}(R)$ indicates that the behavior of the wave's amplitude will depend on its frequency. Let us start by determining this behavior for frequencies $\omega\gg\xi^2$ (or, in other words, wavelengths much smaller than the~$AdS$ radius, $(2\xi)^{-2}$). Given that the bulk of the weight of~$\rho(\alpha)$ is concentrated in the region of~$\alpha\lesssim\cal{O}(1)$, we can safely extend the integration to the infinity and expand the square root in the exponential to obtain:
\begin{equation}
{\mathbf G}^\omega_{(f,f)}(R) \approx \frac{3a}{2\pi^3 }\frac{e^{i\omega R}}{R}\int_0^{\infty}\frac{d\alpha}{\alpha^3} \,\frac{\exp\left[-i\frac{2\xi^4 R}{\omega}\alpha^2 \right]}{J^2_2(\alpha) +Y^2_2(\alpha)} \ .
\end{equation}
For distances $R\ll\omega/\xi^4$, the phase factor varies very slowly and can be set to 1. We obtain then:
\begin{equation}
{\mathbf G}_{(f,f)}^\omega(R) \approx \frac{3a}{2\pi^3 }\frac{e^{i\omega R}}{R}\int_0^{\infty}\frac{d\alpha}{\alpha^3} \,\frac{1}{J^2_2(\alpha) +Y^2_2(\alpha)}=\frac{3a}{16 \pi }\frac{e^{i\omega R}}{R}\ .
\end{equation}
where we have used~\eqref{exact_int}. We therefore recover the usual~$1/R$ dependence of the wave amplitude on the distance to the source. 
 
At very large distances, for $R\gg\omega/\xi^4$, the phase factor varies very rapidly and only the small~$\alpha$'s will contribute significantly to the integral. We can therefore approximate the Bessel functions by the first terms of their series expansion:
$$
{\mathbf G}_{(f,f)}^\omega(R) \approx \frac{3a}{16 \pi }\frac{e^{i\omega R}}{R}\int_0^{\infty}{d\alpha} \,\frac{\alpha}{\alpha^2+2}\exp\left[-i\frac{2\xi^4 R}{\omega}\alpha^2 \right]\ .
$$ 
By successive integration by parts, we can express this integral as a series in $\frac{\omega}{\xi^4 R}$:
\begin{equation}
{\mathbf G}_{(f,f)}^\omega(R) \approx \frac{3a}{16 \pi }\frac{e^{i\omega R}}{R}\left[\frac{\omega}{8\xi^4 R}e^{-i\pi/2}+{\cal O}\left(\left[\frac{\omega}{\xi^4 R}\right]^2\right)\right] \ , 
\end{equation}
that is, the amplitude is proportional to~$1/R^2$, signaling the dissipation of the waves into the extra dimension.

Let us now consider low frequencies: $\omega\ll\xi^2$. For these we can simplify the integrand using the small argument approximation of the Bessel functions:
\begin{align*}
{\mathbf G}_{(f,f)}^\omega(R) &\approx \frac{3a}{32\pi R}\int_0^{\omega/2\xi^2} {d\alpha}\, \alpha \exp\left[i\omega R\sqrt{1-(2\xi^2/\omega)^2\alpha^2}\right]\\
&=\frac{3a\omega}{128 \pi\xi^4}e^{-i\pi/2}\left[\frac{e^{i\omega R}}{R^2}-2\sin\left(\frac{\omega R}{2}\right)\frac{e^{i\omega R/2}}{\omega R^3}\right]\ .
\end{align*} 
It is transparent that in this regime of frequencies, the leaking into the fifth dimension is important at all distances. Dominating behavior at large distances is again of~$1/R^2$ type.

\section{Conclusions}

We have studied the interactions of the scalar perturbations with sources on the brane, both static and oscillating. Our analysis reveals that at distances smaller than the~$AdS$ radius,~$R\ll \xi^{-2}$, the light modes of the continuous spectrum induce a static potential which tends smoothly to the zero gravity limit. Studying the propagation of a signal emitted by an oscillating source on the brane, we have shown that for wave frequencies higher than the inverse of the~$AdS$ radius, the wave amplitude has the usual four-dimensional behavior in a large interval of distances, scaling as the inverse of the distance to the source. The dissipation of the wave into the fifth dimension becomes significant only at very large distances.

We relate these phenomena to a wide resonance, which can be seen as a remnant of the scalar zero mode present in the domain wall setup in the absence of gravity~\cite{shap}. Presence of this resonance ensures a continuous change of the physical quantities when the Planck mass is sent to infinity. Provided that the thickness of the wall is much smaller than the~$AdS$ radius of the space-time, the parameters of this resonance do not depend on details of the domain wall's structure, but only on the geometry of the space-time. 

We did not derive here an explicit low energy effective action for our setup. In the non gravitating limit, we expect it to coincide with the Nambu-Goto action~\cite{sundrum_eff}. The translational zero mode can be then identified with the branon excitation. Our results show that the gravitational backreaction of the brane on the bulk geometry has a strong influence on the appearance of the mode. The effective action for our self-gravitating model together with phenomenological implications of our results and their relation with massive branons are under study and will be discussed elsewhere. 

\acknowledgments{We wish to thank S.~Dubovsky and A.~Neronov for helpful discussions. This work is supported in part by Swiss National Science Foundation.} 

\appendix

\section{\label{app-mu=0_sol}\boldmath $\mu=0$ solutions of the eigenvalue equation of~${\cal{H}}_1$}

In this appendix we present the equations which allow to determine the solutions of the eigenvalue equation of the operator~${\cal{H}}_1$ for~$\mu=0$ and its general solution. In this case, the general solution of \eqref{system} can in fact be determined exactly. 
 
First of all, as suggested by the factorization of ${\cal{H}}_1$, we get two solutions solving:
$$
{\cal A}\begin{pmatrix}h_0 \cr f_0\end{pmatrix}=e^{-{A}/{4}} \begin{pmatrix}\displaystyle \frac{d}{dr}+\frac{1}{4}A' &\displaystyle -\sqrt{\frac{2\kappa}{3}}\Phi' \cr \displaystyle -\sqrt{\frac{2\kappa}{3}}\Phi' & \displaystyle\frac{d}{dr}-\frac{3}{4}A'-\frac{\Phi''}{\Phi'}\end{pmatrix} e^{{3}A/4}\begin{pmatrix}h_0 \cr f_0\end{pmatrix}=0 \ .
$$
The above system of first order equations can be easily decoupled by eliminating one of the variables, giving e.g.: 
$$
\begin{cases}
\displaystyle h_0'' +\left(A'-2\frac{\Phi''}{\Phi'}\right)h_0'+\left(2A''-2A'\frac{\Phi''}{\Phi'}\right) h_0 =0 \ ,\cr
\displaystyle f_0=\sqrt{\frac{3}{2\kappa}}\frac{1}{\Phi'}\left[h_0'+A'h_0\right]\ ,
\end{cases}
$$
whose solutions are:
\begin{align*}
\begin{pmatrix}h_0^{(1)}\\[2.5ex] f_0^{(1)}\end{pmatrix}&=\begin{pmatrix} A'e^{-A}\\[2.5ex] \displaystyle-\smash{\sqrt{\frac{2\kappa}{3}}}\Phi'e^{-A} \end{pmatrix} \ ,\\[1ex]
\begin{pmatrix}h_0^{(2)} \\[2.5ex] f_0^{(2)}\end{pmatrix}&=\begin{pmatrix}\displaystyle 1-A'e^{-A}\smash{\int_0^r dy e^{A(y)}} \\[2.5ex] \displaystyle\smash{\sqrt{\frac{2\kappa}{3}}}\Phi'e^{-A}\smash{\int_0^r dy e^{A(y)}} \end{pmatrix} \ .
\end{align*}
The remaining two solutions can be found by solving an inhomogenous equation 
\begin{equation}
\label{inhom}
{\cal A}\begin{pmatrix}h_0 \cr f_0\end{pmatrix}=\begin{pmatrix}\hhh_0 \cr \fff_0\end{pmatrix} \ ,
\end{equation}
where the inhomogenous term is the solution of 
${\cal A}^\dagger\Bigl(\begin{smallmatrix}\hhh_0 \cr \fff_0\end{smallmatrix}\Bigr)=0$, given by:
$$
\begin{pmatrix}\hhh_0 \\[2.5ex] \fff_0\end{pmatrix}=D_1^{(0)}\begin{pmatrix}\displaystyle e^{-3A/2} \\[2.5ex] \displaystyle \smash{\sqrt{\frac{3}{2\kappa}}e^{-3A/2}\frac{A'}{\Phi'}}\end{pmatrix} +D_2^{(0)}\begin{pmatrix}\displaystyle\smash{ e^{-3A/2}\int_0^r dy e^A}\\[2.5ex] \displaystyle\smash{ -\sqrt{\frac{3}{2\kappa}} \left[e^{-A/2}\frac{1}{\Phi'}- e^{-3A/2}\frac{A'}{\Phi'}\int_0^r dy e^A\right]}\end{pmatrix} \ .
$$
The two additional solutions coming from~\eqref{inhom} are of the following form:
\begin{align*}
\begin{pmatrix}h_0^{(3)} \\[2.5ex] f_0^{(3)}\end{pmatrix}&=
\begin{pmatrix}\displaystyle \frac{3}{2\kappa}\!\left[ -e^{-2A}\frac{A'}{\Phi'^2} -\int_0^r dr_1e^{-2A}\frac{A'}{\Phi'^3}\left[ 2\Phi''+A'\Phi'\right]
+\right. \hfill\cr 
\hfill\displaystyle \left.\ \quad+ A'e^{-A}\!\int_0^r dr_1e^{A}\int_0^{r_1}dr_2e^{-2A}\frac{A'}{\Phi'^3}\left[ 2\Phi''+A'\Phi'\right]\right]\\[2.5ex]
 \displaystyle\smash{-\sqrt{\frac{3}{2\kappa}}\Phi'e^{-A}\int_0^r dr_1 e^{A}\int_0^{r_1} dr_2 e^{-2A}}\frac{A'}{\Phi'^3}\left[2\Phi''+A'\Phi'\right]\end{pmatrix}\ , \\[1ex]
\begin{pmatrix}h_0^{(4)}\\[3ex] f_0^{(4)}\end{pmatrix}&=
\begin{pmatrix}\displaystyle \frac{3}{2\kappa}\!\left[ -e^{-2A}\frac{A'}{\Phi'^2}\int_0^r dr_1e^A + A'e^{-A}\!\int_0^r dr_1\frac{1}{\Phi'^2}-{}\right.
\hfill\\[1.5ex]
\displaystyle
\ \quad-\int_0^r dr_1e^{-2A}\frac{A'}{\Phi'^3}\left[ 2\Phi''+A'\Phi'\right]\int_0^{r_1} dr_2 e^A +{} 
\hfill\\[1.5ex]
\displaystyle \left.\ \quad+ A'e^{-A}\!\int_0^r dr_1e^{A}\!\int_0^{r_1}dr_2e^{-2A}\frac{A'}{\Phi'^3}\left[ 2\Phi''+A'\Phi'\right]\!\int_0^{r_2} dr_3 e^A \right]\hfill\\[3ex]
\displaystyle -\sqrt{\frac{3}{2\kappa}}\Phi'e^{-A}\!\left[\int_0^r \frac{dr_1}{\Phi'^2}
+2\!\int_0^r \!dr_1 e^{A}\!\!\int_0^{r_1} \!dr_2 e^{-2A}\frac{A'}{\Phi'^3}\!\left[2\Phi''+A'\Phi'\right]\!\int_0^{r_2}\! dr_3 e^{A}\right]\end{pmatrix}\ .\!\!
\end{align*}
None of these solutions is normalizable.

\section{\label{app-connection}\boldmath Connection between the spectra of~${\cal{H}}_1$ and~${\cal{H}}_2$}

As long as~$\beta\neq 0$, the operators~${\cal{H}}_1$ and~${\cal{H}}_2$ have both purely continuous spectra. In this appendix we will illustrate in some detail how the eigenfunctions of the two operators are related.

Suppose that we have an orthogonal basis of the eigenfunctions of the operator~${\cal H}_2$ for some boundary conditions. As the partner system is decoupled, the natural basis is composed of 
$$
\tilde{\Psi}_{\mu(\hh)}=\begin{pmatrix}\hh_\mu \cr 0\end{pmatrix}\qquad \mbox{and} \qquad \tilde{\Psi}_{\mu(\ff)}=\begin{pmatrix}0 \cr \ff_\mu\end{pmatrix}\ ,
$$ 
where, trivially, $\tilde\Psi_{\mu(\hh)}$-type solution vectors are orthogonal to $\tilde\Psi_{\mu(\ff)}$-type ones. The basis of the eigenfunctions for the coupled system can be obtained from these using~\eqref{recipe}:
$$
\Psi_{\mu(\hh)}=\frac{1}{\mu}{\cal A}^{\dagger}\begin{pmatrix}\hh_\mu \cr 0\end{pmatrix} \qquad \mbox{and} \qquad \Psi_{\mu(\ff)}=\frac{1}{\mu}{\cal A}^{\dagger}\begin{pmatrix}0 \cr \ff_\mu\end{pmatrix} \ , 
$$
where again $\Psi_{\mu(\hh)}$-type solution vectors are orthogonal to $\Psi_{\mu(\ff)}$-type ones. Indeed, we have 
\begin{align*}
\int_{-\infty}^\infty dr \,e^{A(r)}\Psi_{\mu(\hh)}^T\Psi^{}_{\mu'(\ff)}&=\frac{1}{\mu^2}\int_{-\infty}^\infty dr \,e^{A(r)}\left[{\cal A}^{\dagger}\tilde{\Psi}_{\mu(\hh)}\right]^T\left[{\cal A}^{\dagger}\tilde{\Psi}_{\mu'(\ff)}\right] \\
&=\frac{1}{\mu^2}\int_{-\infty}^\infty dr \,e^{A(r)}\left[{\cal A}{\cal A}^{\dagger}\tilde{\Psi}_{\mu(\hh)}\right]^T\tilde{\Psi}_{\mu'(\ff)}\\
&=\int_{-\infty}^\infty dr \,e^{A(r)}\tilde{\Psi}_{\mu(\hh)}^T\tilde{\Psi}^{}_{\mu'(\ff)}=0\ .
\end{align*}
Moreover, the following relations are satisfied:
\begin{align*}
\int_{-\infty}^\infty dr \,e^{A(r)}\Psi_{\mu(\hh)}^T\Psi^{}_{\mu'(\hh)}&=\int_{-\infty}^\infty dr \,e^{A(r)}\tilde{\Psi}_{\mu(\hh)}^T\tilde{\Psi}^{}_{\mu'(\hh)}=\int_{-\infty}^\infty dr \,e^{A(r)}\hh_\mu\hh_{\mu'} \ ,\\
\int_{-\infty}^\infty dr \,e^{A(r)}\Psi_{\mu(\ff)}^T\Psi^{}_{\mu'(\ff)}&=\int_{-\infty}^\infty dr \,e^{A(r)}\tilde{\Psi}_{\mu(\ff)}^T\tilde{\Psi}^{}_{\mu'(\ff)}=\int_{-\infty}^\infty dr \,e^{A(r)}\ff_\mu\ff_{\mu'} \ .
\end{align*}
Consequently, normalizing the basis solutions of the partner system using:
\begin{align}
\label{norm_partner_h}
\int_{-\infty}^\infty dr \,e^{A(r)}\hh_\mu(r)\hh_{\mu'}(r)&= \delta(\mu-\mu')\ ,\\
\label{norm_partner_f}
\int_{-\infty}^\infty dr \,e^{A(r)}\ff_\mu(r)\ff_{\mu'}(r)&= \delta(\mu-\mu') \ ,
\end{align}
we obtain an orthonormal basis of eigenfunctions ${\cal H}_1$:
\begin{equation}
\label{ortho-Psi}
\int_{-\infty}^\infty dr \,e^{A(r)}\Psi_{\mu(\alpha)}^T\Psi^{}_{\mu'(\beta)}=\delta(\mu-\mu')\delta_{\alpha\beta}\ ,
\end{equation}
where $\alpha,\beta\in\{\hh,\ff\}$. The solutions~$\Psi_{\mu(\alpha)}$ satisfy the following completeness relation:
\begin{equation}
\label{compl-Psi}
\sum_\alpha\int_{0}^\infty d\mu \,\Psi^{}_{\mu(\alpha)}(r)\Psi_{\mu(\alpha)}^T(r')=\,e^{-A(r)}\delta(r-r')1\!\!1_2\ .
\end{equation}

\section{\label{app-eigenfunctions}\boldmath Eigenfunctions of the partner operator~${\cal{H}}_2$}

This appendix contains a detailed derivation of the eigenfunctions of the partner operator
$$
{\cal H}_2=e^A\mbox{\footnotesize$
\begin{pmatrix}\displaystyle -\frac{d^2}{dr^2}-2A'\frac{d}{dr}
-\left(\frac32A''+\frac34A'^2\right)& 0 
\cr
 0 & \displaystyle \!\!\!-\frac{d^2}{dr^2}-2A'\frac{d}{dr}-\left[\frac{\Phi'''}{\Phi'}-2\!\left(\frac{\Phi''}{\Phi'}\right)^{\!2}-A'\frac{\Phi''}{\Phi'}+\frac52A''+\frac34A'^2\right]\end{pmatrix}$} \ .\!\!\!\!\!
$$
For $\mu=0$ the general solution of the system~\eqref{partner} can be calculated exactly:
\begin{align*}
\begin{pmatrix}\hh_0 \cr \ff_0\end{pmatrix}&=D_1^{(0)}\begin{pmatrix}\displaystyle e^{-3A/2} \cr 0\end{pmatrix} +D_2^{(0)}\begin{pmatrix}\displaystyle e^{-3A/2}\int_0^r dy e^A \cr 0\end{pmatrix} +\\
&\quad +D_3^{(0)}\begin{pmatrix} 0 \cr\displaystyle e^{-3A/2}\frac{A'}{\Phi'}\end{pmatrix}+D_4^{(0)}\begin{pmatrix} 0 \cr \displaystyle e^{-A/2}\frac{1}{\Phi'}- e^{-3A/2}\frac{A'}{\Phi'}\int_0^r dy e^A\end{pmatrix} \ .
\end{align*}
There are two combinations of the four basis solutions which are solutions of \mbox{${\cal A}^\dagger\Bigl(\begin{smallmatrix}\hhh_0 \cr \fff_0\end{smallmatrix}\Bigr)=0$}. These correspond to~\mbox{$D_3^{(0)}=\sqrt{\frac{3}{2\kappa}}D_1^{(0)}$} and~$D_4^{(0)}=-\sqrt{\frac{3}{2\kappa}}D_2^{(0)}$.

For non-zero~$\mu$ it is impossible to find exact solutions. However, one can find approximate solutions of the system using the thin wall limit. The partner system being decoupled, our task is reduced to finding solutions of two second order equations. 

\noindent $\bullet$ Let us start by determining the solutions of the first of the equations~\eqref{partner}:
\begin{equation}
\label{eqn_hh}
-\hh'' - 2A'\hh'-\left(\frac{3}{2}A''+\frac{3}{4}A'^2\right)\hh =\mu^2 e^{-A}\hh \ .
\end{equation}
The exponent of the warp factor~$A(r)$ is given by Eqn.~\eqref{A(r)}. In the thin wall limit~\eqref{tw}, we have
\begin{align*}
A & \approx - 4 \xi^2 |r| \ ,\\
A'& \approx - 4\xi^2\mathop{\rm sign}(r) \ ,\\
A''& \approx -8\xi^2 \delta(r) \ .
\end{align*}
Consequently, Eqn.~\eqref{eqn_hh} becomes :
\begin{equation}
\label{eqn_hh_tw}
-\hh'' + 8 \xi^2\mathop{\rm sign}(r)\hh'+\left(12\xi^2\delta(r)- 12\xi^4\right)\hh ={\mu}^2 e^{4\xi^2|r|}\hh \ ,
\end{equation}
where $\xi^2\equiv a\beta$. This equation can be solved for arbitrary~$\mu$ and its solutions are:
$$
\hh_\mu(r)= 
\begin{cases}\displaystyle e^{4\xi^2 r}\left[B_1 J_1\left(\frac{\mu}{2\xi^2}e^{2\xi^2 r}\right)+B_2 Y_1\left(\frac{\mu}{2\xi^2}e^{2\xi^2 r}\right)\right] & r>0 \ ,\\[1.7ex] \displaystyle
e^{-4\xi^2 r}\left[B_3 J_1\left(\frac{\mu}{2\xi^2}e^{-2\xi^2 r}\right)+B_4 Y_1\left(\frac{\mu}{2\xi^2}e^{-2\xi^2 r}\right)\right] & r<0 \ .
\end{cases}
$$
Eqn.~\eqref{eqn_hh} being symmetric under reflection~$r\to-r$, for any given~$\mu$ there exists an even and an odd solution. Since they are, of course, linearly independent, the general solution can be written as a linear combination of them. We can therefore consider solutions with definite parity. 
For an even solution, we must have $B_3=B_1$ and $B_4=B_2$ and it can thus be written as:
\begin{equation}
\label{hh_even_tw}
\hh_\mu^{(e)}(r)= e^{4\xi^2 |r|}\left[B_1^{(e)} J_1\left(\frac{\mu}{2\xi^2}e^{2\xi^2 |r|}\right)+B_2^{(e)} Y_1\left(\frac{\mu}{2\xi^2}e^{2\xi^2 |r|}\right)\right] \ .
\end{equation}
For~$\hh^{(e)}(r)$ to be a solution of~\eqref{eqn_hh_tw}, it is necessary that the coefficients $B_1^{(e)}$ and $B_2^{(e)}$ satisfy the following relation (imposed by the jump of the derivative at~$r=0$): 
\begin{equation}
\label{hh_even_cond}
B_2^{(e)}=-B_1^{(e)}\frac{J_1\left(\frac{\mu}{2\xi^2}\right)-\frac{\mu}{4\xi^2}J_0\left(\frac{\mu}{2\xi^2}\right)}{Y_1\left(\frac{\mu}{2\xi^2}\right)-\frac{\mu}{4\xi^2}Y_0\left(\frac{\mu}{2\xi^2}\right)}=-B_1^{(e)}\frac{J_2\left(\frac{\mu}{2\xi^2}\right)}{Y_2\left(\frac{\mu}{2\xi^2}\right)} \quad .
\end{equation}
Imposing the normalization condition~\eqref{norm_partner_h}, we fix the remaining constant:
\begin{equation}
\label{B_1_e}
\left[ B_1^{(e)}(\mu)\right]^2=\frac{\mu}{4\xi^2}\left[1+\left(\frac{J_2\left(\frac{\mu}{2\xi^2}\right)}{Y_2\left(\frac{\mu}{2\xi^2}\right)}\right)^{\!\!\!2}\ \right]^{-1} \ .
\end{equation}
For an odd solution, we must have $B_3=-B_1$ and $B_4=-B_2$. Thus, it can be written as:
\begin{equation}
\label{hh_odd_tw}
\hh_\mu^{(o)}(r)= \mathop{\rm sign}(r)\,e^{4\xi^2 |r|}\left[B_1^{(o)} J_1\left(\frac{\mu}{2\xi^2}e^{2\xi^2 |r|}\right)+B_2^{(o)} Y_1\left(\frac{\mu}{2\xi^2}e^{2\xi^2 |r|}\right)\right] \ .
\end{equation}
The coefficients $B_1^{(o)}$ and $B_2^{(o)}$ must satisfy the following relation (imposed to avoid $\delta'$-like singularities):
\begin{equation}
\label{hh_odd_cond}
B_2^{(o)}=-B_1^{(o)}\frac{J_1\left(\frac{\mu}{2\xi^2}\right)}{Y_1\left(\frac{\mu}{2\xi^2}\right)} \ .
\end{equation}
Imposing the normalization condition~\eqref{norm_partner_h}, we find:
\begin{equation}
\label{B_1_o}
\left[ B_1^{(o)}(\mu)\right]^2=\frac{\mu}{4\xi^2}\left[1+\left(\frac{J_1\left(\frac{\mu}{2\xi^2}\right)}{Y_1\left(\frac{\mu}{2\xi^2}\right)}\right)^{\!\!\!2}\ \right]^{-1} \ .
\end{equation}

\noindent $\bullet$ Let us now consider the second of the equations~\eqref{system}:
\begin{equation}
\label{eqn_ff}
-\ff '' - 2A'\ff'-\left[\frac{\Phi'''}{\Phi'}-2\left(\frac{\Phi''}{\Phi'}\right)^2-A'\frac{\Phi''}{\Phi'}+\frac{5}{2}A''+\frac{3}{4}A'^2\right]\ff =\mu^2 e^{-A}\ff \ .
\end{equation}
In the thin wall limit, for $|r|\gg1/a$, this equation becomes:
\begin{equation}
\label{eqn_ff_tw}
-\ff '' +8\xi^2\mathop{\rm sign}(r)\ff'+\left(4a^2+8\xi^2a-12\xi^4\right)\ff=\mu^2 e^{4\xi^2 |r|}\ff\ ,
\end{equation} 
the solutions of which are:
\begin{equation}
\label{ff_sols}
\ff_\mu= 
\begin{cases}\displaystyle e^{4\xi^2 r}\left[D_1 J_\nu\left(\frac{\mu}{2\xi^2}e^{2\xi^2 r}\right)+D_2 J_{-\nu}\left(\frac{\mu}{2\xi^2}e^{2\xi^2 r}\right)\right] & r>0 \ , \\[1.5ex] \displaystyle
e^{-4\xi^2 r}\left[D_3 J_\nu\left(\frac{\mu}{2\xi^2}e^{-2\xi^2 r}\right)+D_4 J_{-\nu}\left(\frac{\mu}{2\xi^2}e^{-2\xi^2 r}\right)\right] & r<0 \ ,
\end{cases}
\end{equation}
where $\nu=1/\beta+1$. We cannot extend them towards~$r=0$ as simply as it could be done for the solutions of~Eqn.~\eqref{eqn_hh_tw}. However, it is still possible to find (approximate) global solutions of~\eqref{eqn_ff_tw} for small~$\mu$ matching~\eqref{ff_sols} to the exact~$\mu=0$ solutions.

Again, given the reflection symmetry of~\eqref{eqn_ff} we can consider even and odd solutions separately. We will limit ourselves to determining the odd solutions, as only these are needed in the calculations performed in sections~\ref{sect-static} and~\ref{sect-periodic} . The even solutions can be found in an entirely analogous manner. 
The odd~$\mu=0$ solution is:
$$
\ff_0^{(o)}=e^{-3A/2}\frac{A'}{\Phi'}\ .
$$
For small $\mu$, the global odd solution of~\eqref{eqn_ff_tw} can therefore be approximated by:
$$
\ff_\mu^{(o)}(r)\approx 
\begin{cases}\displaystyle C^{(o)}\,e^{-3A/2}\frac{A'}{\Phi'} & \displaystyle |r|<d \ , \\[1ex] \displaystyle \mathop{\rm sign}(r)\,e^{4\xi^2 |r|}\left[D_1^{(o)}J_\nu\left(\frac{\mu}{2\xi^2}e^{2\xi^2 |r|}\right)+D_2^{(o)}J_{-\nu}\left(\frac{\mu}{2\xi^2}e^{2\xi^2 |r|}\right)\right] & |r|\gg\displaystyle\frac{1}{a} \ ,
\end{cases}
$$
where $d$ must be such that the~$\mu^2$ (and higher) corrections to the $\mu=0$ term are negligible. We must have $d\lesssim \displaystyle (2\xi^2)^{-1}\ln(2\xi^2\nu\mu^{-1})$. In order to be able to do the matching, we can safely take $d\sim{\cal O}(\xi^{-2})$. Then our solution for~$f_\mu$ is valid for $\mu\lesssim \nu \xi^2\sim a$. For $|r|\gg 1/a$ the exact solution~$\ff_0^{(o)}$ can be expanded in series in~$e^{-2a|r|}$:
$$
\ff_0^{(o)}=e^{-3A/2}\frac{A'}{\Phi'} \approx-\mathop{\rm sign}(r)\frac{\xi^2}{va}\, e^{2a|r|+6\xi^2|r|}\left(1+2e^{-2a|r|}+{\cal{O}}(e^{-4a|r|})\right) \\
$$
and in the overlap region~$a^{-1}\ll|r|<d$ we have 
\begin{align*}
e^{4\xi^2 |r|}J_\nu\left(\frac{\mu}{2\xi^2}e^{2\xi^2 |r|}\right)& \approx \frac{1}{2\pi\nu}\left(\frac{e\,\mu}{4\nu\xi^2}\right)^{\nu}e^{2a|r|+6\xi^2 |r|}\ , \\
e^{4\xi^2 |r|}J_{-\nu}\left(\frac{\mu}{2\xi^2}e^{2\xi^2 |r|}\right)& \approx \frac{1}{2\pi\nu}\left(\frac{e\,\mu}{4\nu\xi^2}\right)^{-\nu} e^{-2a|r|+2\xi^2 |r|}\ .
\end{align*}
Comparing the above three expressions, we obtain (at the lowest order in~$\mu$)~$D_2^{(o)}$=0 and 
$$
C^{(o)}\approx-\frac{va}{2\pi\nu\xi^2}\left(\frac{e\,\mu}{4\nu\xi^2}\right)^{\nu} D_1^{(o)}\approx -\frac{v}{2\pi}\left(\frac{e\mu}{4\nu\xi^2}\right)^{\nu} D_1^{(o)} \ .
$$
Therefore, in the first approximation, the solution is 
\begin{equation}
\label{ff_odd_tw}
\ff^{(o)}(r)\approx D_1^{(o)}
\begin{cases}\displaystyle -\frac{v}{2\pi}\left(\frac{e\mu}{4\nu\xi^2}\right)^{\nu}\,e^{-3A/2}\frac{A'}{\Phi'} & |r|<d\ , \\[1.5ex] \displaystyle 
\mathop{\rm sign}(r)\, e^{4\xi^2 |r|}J_\nu\left(\frac{\mu}{2\xi^2}e^{2\xi^2 |r|}\right) & |r|\gg\displaystyle\frac{1}{a}\ .
\end{cases}
\end{equation}
Imposing the normalization condition~\eqref{norm_partner_f} we fix the remaining constant~$D_1^{(o)}$;
\begin{equation}
\label{D_1_o}
\left[D_1^{(o)}(\mu)\right]^2=\frac{\mu}{4\xi^2}\ ,
\end{equation}
which completes the determination of~$\ff^{(o)}(r)$. 

\section{\label{app-static}Determination of the static potential}

In this appendix, we present the details of the calculation of the potential between two static sources of~$\phi$. The field produced by the source $\underline{J}$ is given by:
\begin{equation}
\label{field}
\Psi(x,r)=\int d^4x'dr' e^{2A(r')}{\mathbf G}(x,r;x',r')\underline{J}(x',r')\ ,
\end{equation}
where ${\mathbf G}(x,r;x',r')$ is the (matrix) retarded Green's function of the coupled system~\eqref{system_chi_phi}, satisfying
\begin{eqnarray*}
\left[\begin{pmatrix}e^{-A}\eta^{\mu\nu}\partial_\mu\partial_\nu& 0 \cr 0&e^{-A}\eta^{\mu\nu}\partial_\mu\partial_\nu\end{pmatrix}+ e^{-A} {\cal H}_1\right] 
{\mathbf G}(x,r;x',r')=\delta^{(4)}(x-x')\delta(r-r')e^{-2A(r)} 1\!\!1_2 \ .
\end{eqnarray*} 
The Green's function ${\mathbf G}(x,r;x',r')$ can be written as a Fourier transform of the Green's function~${\mathbf G}^k_{(5)}(r,r')$ of our one-dimensional matrix Sturm-Liouville problem~\eqref{system}:
\begin{equation}
\label{green_def}
({\cal H}_1-k^21\!\!1_2) {\mathbf G}^k_{(5)}(r,r')=\delta(r-r')e^{-A(r)} 1\!\!1_2 \ .
\end{equation}
As per usual, ${\mathbf G}^k_{(5)}(r,r')$ can be expanded in the basis of eigenfunctions~$\Psi_{\mu(\alpha)}$, $\alpha\in\{\hh,\ff\}$ of the operator~${\cal H}_1$, which allows us to write the following expression for the retarded Green's function:
\begin{align}
{\mathbf G}(x,r;x',r')&=\int \frac{d^4k}{(2\pi)^4}e^{ik_\mu(x^\mu -x'^\mu)} {\mathbf G}^k_{(5)}(r,r')\nonumber\\
&=\int \frac{d^4k}{(2\pi)^4}e^{ik_\mu(x^\mu -x'^\mu)}\int_0^\infty d\mu\, \sum_\alpha \,\frac{\Psi_{\mu(\alpha)}(r)\Psi_{\mu(\alpha)}^T(r')}{\mu^2-k^2-2i\varepsilon k_0}\quad .
\label{green_total}\end{align}
If the source $\underline{J}$ is static, we can perform in~\eqref{field} the integration over time on the Green's function and obtain 
\begin{align*}
\Psi(x,r)&=\int d^3x'dr' e^{2A(r')}\left(\int_{-\infty}^{\infty}dt' {\mathbf G}(x,r;x',r')\right) \underline{J}(\vec{x}',r') \\
&= \frac{1}{4\pi}\int d^3x'dr'\,e^{2A(r')}\int_0^\infty d\mu \frac{e^{-\mu|\vec{x}-\vec{x}'|}}{|\vec{x}-\vec{x}'|}\sum_\alpha \,{\Psi_{\mu(\alpha)}(r)\Psi_{\mu(\alpha)}^T(r')}\underline{J}(\vec{x}',r')\ .
\end{align*}
The static potential is the interaction energy of a source $\underline{J}_{(1)}=\Bigl(\begin{smallmatrix}0\cr J^{(1)}_f\end{smallmatrix}\Bigr)$ with the field $\Psi_{(2)}$ produced by~$\underline{J}_{(2)}=\Bigl(\begin{smallmatrix}0\cr J^{(2)}_f\end{smallmatrix}\Bigr)$, that is:
\begin{align*}
V&=\int d^3x \,dr \,e^{2A(r)}\underline{J}_{(1)}^T(\vec{x},r)\Psi_{(2)}(\vec{x},r)\\ &= \int d^3x \,dr\, e^{2A(r)}\int d^4x'dr'\,e^{2A(r')}\underline{J}_{(1)}^T(\vec{x},r) {\mathbf G}(x,r;x',r')\underline{J}_{(2)}(x',r') 
\\
&=\frac{1}{4\pi} \int d^3x\, dr \,e^{2A(r)} \int d^3x'dr' \,e^{2A(r')}\times\\
&\ \qquad\qquad\times\int_0^\infty d\mu \frac{e^{-\mu|\vec{x}-\vec{x}'|}}{|\vec{x}-\vec{x}'|}\sum_\alpha \,\underline{J}_{(1)}^T(\vec{x},r){\Psi_{\mu(\alpha)}(r)\Psi_{\mu(\alpha)}^T(r')}\underline{J}_{(2)}(\vec{x}',r')\ .
\end{align*}
Suppose that the sources are pointlike and situated at $r=r'=0$, that is
\begin{align*}
\underline{J}_{(1)}(\vec{x},r)&=\underline{\tilde{J}}_{(1)}\,\delta^3(\vec{x}-\vec{x}_1)\delta(r) \ ,\\
\underline{J}_{(2)}(\vec{x},r)&=\underline{\tilde{J}}_{(2)}\,\delta^3(\vec{x}-\vec{x}_2)\delta(r) \ .
\end{align*}
Then, the static potential~$V$ becomes:
\begin{align*}
V(|\vec{x}_1-\vec{x}_2|)&=\frac{1}{4\pi}\int_0^\infty d\mu \frac{e^{-\mu|\vec{x}_1-\vec{x}_2|}}{|\vec{x}_1-\vec{x}_2|}\sum_\alpha\underline{\tilde{J}}_{(1)}^T\Psi_{\mu(\alpha)}(0)\Psi^T_{\mu(\alpha)}(0)\underline{\tilde{J}}_{(2)} \nonumber \\
&=\frac{1}{4\pi}{\tilde{J}}_{(1),f}{\tilde{J}}_{(2),f}\int_0^\infty d\mu \frac{e^{-\mu|\vec{x}_1-\vec{x}_2|}}{|\vec{x}_1-\vec{x}_2|} \sum_\alpha \left[f_{\mu(\alpha)}(0)\right]^2 \ ,
\end{align*}
where in the last line we have used the fact that both sources are the type $\Bigl(\begin{smallmatrix}0\cr J_f\end{smallmatrix}\Bigr)$.

\section{\label{app-sol-coupled}Solutions of the coupled system}

In this appendix we determine the solutions of the coupled system~${\cal H}_1$, using its connection with its (decoupled) partner system~${\cal H}_2$ via the relation~\eqref{recipe}.

Applying ${\cal A}^{\dagger}$ on the even solution of the eqn.~\eqref{eqn_hh}, given by~\eqref{hh_even_tw}-\eqref{B_1_e}, we obtain:
\begin{align*}
\Psi_{\mu(\hh)}&=\begin{pmatrix}h_{\mu(\hh)}\\[1ex] f_{\mu(\hh)}\end{pmatrix}=\frac{1}{\mu}{\cal A}^{\dagger}\begin{pmatrix}\hh_\mu^{(e)} \\[1ex] 0\end{pmatrix}=\frac{1}{\mu}\begin{pmatrix}e^{A/2}\left[-\hh_\mu^{(e)\prime}-\frac12A'\hh_\mu^{(e)}\right] \\[1ex] -\sqrt{\frac{2\kappa}{3}}e^{A/2}\Phi'\hh_\mu^{(e)} \end{pmatrix} \\[1ex]&\approx \frac{1}{\mu}\begin{pmatrix}e^{-2\xi^2|r|}\left[-\hh_\mu^{(e)\prime}+2\xi^2\mathop{\rm sign}(r)\hh_\mu^{(e)}\right] \\[1ex] -\sqrt{\frac{2\kappa v^2}{3}}a\mathop{\rm sech}^2(ar)e^{-2\xi^2|r|}\hh_\mu^{(e)} \end{pmatrix} \\[1ex]
&\approx \frac{B_1^{(e)}(\mu)}{\mu}\begin{pmatrix}\displaystyle - \mu\mathop{\rm sign}(r)\,e^{4\xi^2 |r|}\,\left[J_0\left(\frac{\mu}{2\xi^2}e^{2\xi^2 |r|}\right)-\frac{J_2\left(\frac{\mu}{2\xi^2}\right)}{Y_2\left(\frac{\mu}{2\xi^2}\right)}Y_0\left(\frac{\mu}{2\xi^2}e^{2\xi^2 |r|}\right)\right]\cr\displaystyle - 2\sqrt{6\beta}\ \left[J_1\left(\frac{\mu}{2\xi^2}\right)-\frac{J_2\left(\frac{\mu}{2\xi^2}\right)}{Y_2\left(\frac{\mu}{2\xi^2}\right)}Y_1\left(\frac{\mu}{2\xi^2}\right)\right]\delta(r)\end{pmatrix}\ .
\end{align*}
We have therefore:
\begin{align*}
[f_{\mu(\hh)}(0)]^2&= -\mu^{-2}\frac{6\beta}{v^2}\left[\Phi'(0) \hh_\mu^{(e)}\right]^2= \frac{6\beta a^2}{\mu^2} \left[B_1^{(e)}(\mu)\right]^2\left[J_1\left(\frac{\mu}{2\xi^2}\right)-\frac{J_2\left(\frac{\mu}{2\xi^2}\right)}{Y_2\left(\frac{\mu}{2\xi^2}\right)}Y_1\left(\frac{\mu}{2\xi^2}\right)\right]^2 \\
&=\frac{3}{\pi^2}\frac{1}{\beta}\left(\frac{2\xi^2}{\mu}\right)^3\left[J_2^2\left(\frac{\mu}{2\xi^2}\right)+Y_2^2\left(\frac{\mu}{2\xi^2}\right)\ \right]^{-1}\ ,
\end{align*}
where we have used~\eqref{B_1_e}.

Applying ${\cal A}^{\dagger}$ on the odd solution of the eqn.~\eqref{eqn_ff}, given by~\eqref{ff_odd_tw}--\eqref{D_1_o} we obtain:
$$
\Psi_{\mu(\ff)}=\begin{pmatrix}h_{\mu(\ff)}\cr f_{\mu(\ff)}\end{pmatrix}=\frac{1}{\mu}{\cal A}^{\dagger}\begin{pmatrix}0 \cr \ff_\mu^{(o)}\end{pmatrix}=\frac{1}{\mu}\begin{pmatrix}-\sqrt{\frac{2\kappa}{3}}e^{A/2}\Phi'\ff_\mu^{(o)} \cr e^{A/2}\left[-\ff_\mu^{(o)\prime}-\left(\frac32A'+\frac{\Phi''}{\Phi'}\right)\ff_\mu^{(o)}\right]\end{pmatrix} \ ,
$$
which for $|r|<d$ gives us:
\begin{equation}
\label{f_ff_close}
\frac{1}{\mu}{\cal A}^{\dagger}\begin{pmatrix}0 \cr \ff_\mu^{(o)}\end{pmatrix}=\frac{\sqrt{6\beta}}{2\pi}\left(\frac{e\mu}{4\xi^2\nu}\right)^{\nu} \frac{D_1^{(o)}(\mu)}{\mu}\begin{pmatrix}A'e^{-A} \cr -\sqrt{6\beta}/v\Phi'e^{-A}\end{pmatrix} \ ,
\end{equation}
while for $r\gg1/a$ we have: 
\begin{align*}
\frac{1}{\mu}{\cal A}^{\dagger}\begin{pmatrix}0 \cr \ff_\mu^{(o)}\end{pmatrix}&\approx \frac{D_1^{(o)}(\mu)}{\mu} \begin{pmatrix}-4a \sqrt{6\beta} e^{-2a|r|+2\xi^2|r|}\, \mathop{\rm sign}(r) \, J_\nu\left(\frac{\mu}{2\xi^2}e^{2\xi^2 |r|}\right) \cr \mu e^{4\xi^2 |r|}J_{\nu+1}\left(\frac{\mu}{2\xi^2}e^{2\xi^2 |r|}\right) \end{pmatrix} \ .
\end{align*}
The expression~\eqref{f_ff_close} gives us:
$$
\left[f_{\mu(\ff)}(0)\right]^2\approx\left(\frac{6\beta}{2\pi}\right)^2\left(\frac{e\mu}{4\xi^2\nu}\right)^{2\nu}\left( \frac{\Phi'(0)}{v}\right)^2\left[\frac{D_1^{(o)}(\mu)}{\mu}\right]^2 \approx \frac{9\nu}{16 e} \left(\frac{e\mu}{4\xi^2\nu}\right)^{2\nu-1} \ ,
$$
where we have used~\eqref{D_1_o}.

\providecommand{\href}[2]{#2}\begingroup\raggedright\endgroup
\end{document}